\documentclass[twocolumn,floats,floatfix,amssymb,nofootinbib,prd,superscriptaddress,aps]{revtex4-1}
\usepackage{graphicx,amssymb,amsmath,amsthm,amsfonts,epsfig}
\usepackage{subcaption}
\usepackage[linktocpage,colorlinks=true,citecolor=OliveGreen,linkcolor=Maroon,urlcolor=Maroon]{hyperref}
\usepackage[usenames]{color}
\usepackage{epstopdf}
\usepackage{bm}
\usepackage{dcolumn}
\usepackage[utf8]{inputenc}
\usepackage{latexsym}
\usepackage{rotating}
\usepackage{hyperref}
\usepackage{tabularx}
\usepackage{braket}
\usepackage{color}
\usepackage{esvect}
\usepackage{amsmath}
\usepackage{enumerate}
\usepackage{footmisc}
\usepackage{array}
\usepackage{tensor}
\usepackage{mathtools}
\usepackage{url}
\usepackage[dvipsnames]{xcolor}
\usepackage{comment}
\usepackage{tikz}
\usetikzlibrary{
    arrows.meta,
    shapes.geometric,
    shapes.misc,
    positioning,
    calc}
\usepackage{multirow}
\usepackage{nicematrix}
\usepackage{mathrsfs}
\usepackage{float}
\usepackage{ragged2e}
\usepackage[normalem]{ulem}

\def\be{\begin{equation}}
\def\ee{\end{equation}}
\def\bea{\begin{eqnarray}}
\def\eea{\end{eqnarray}}
\def\beq{\begin{eqnarray}}
\def\eeq{\end{eqnarray}}
\newcommand{\ba}{\begin{align}}
\newcommand{\ea}{\end{align}}

\definecolor{codegray}{gray}{0.95}
\definecolor{keywordcolor}{rgb}{0.26, 0.32, 0.66}
\definecolor{commentcolor}{rgb}{0.0, 0.5, 0.0}

\newcommand{\vc}[1]{{\textcolor{blue}{\sf{[Vitor: #1]}} }}

\newcommand{\fs}[1]{{\textcolor{cyan}{\sf{[Filipe: #1]}} }}
\definecolor{cornellGreen}{HTML}{6EB43F}
\definecolor{cornellRed}{HTML}{B31B1B}

\begin{document}

\title{Wave optical imaging of an oscillating electric dipole orbiting a black hole}
\author{Filipe Nazaré}
\affiliation{Center of Gravity, Niels Bohr Institute, Blegdamsvej 17, 2100 Copenhagen, Denmark}
\affiliation{CENTRA, Departamento de F\'{\i}sica, Instituto Superior T\'ecnico -- IST, Universidade de Lisboa -- UL,
Avenida Rovisco Pais 1, 1049-001 Lisboa, Portugal}

\author{João S. Santos}
\affiliation{Center of Gravity, Niels Bohr Institute, Blegdamsvej 17, 2100 Copenhagen, Denmark}
\affiliation{CENTRA, Departamento de F\'{\i}sica, Instituto Superior T\'ecnico -- IST, Universidade de Lisboa -- UL,
Avenida Rovisco Pais 1, 1049-001 Lisboa, Portugal}

\author{Vitor Cardoso}
\affiliation{Center of Gravity, Niels Bohr Institute, Blegdamsvej 17, 2100 Copenhagen, Denmark}
\affiliation{CENTRA, Departamento de F\'{\i}sica, Instituto Superior T\'ecnico -- IST, Universidade de Lisboa -- UL,
Avenida Rovisco Pais 1, 1049-001 Lisboa, Portugal}

\author{José Natário}
\affiliation{CAMGSD, Departamento de Matemática, Instituto Superior Técnico – IST,
Universidade de Lisboa – UL, Avenida Rovisco Pais 1, 1049-001 Lisboa, Portugal}

\begin{abstract}
We study the electromagnetic radiation and wave-optical imaging of an oscillating electric dipole orbiting a Kerr black hole. We derive the effective $4$-current associated with a pointlike oscillating electric dipole in curved spacetime, and use black hole perturbation theory to compute the resulting radiation field at future null infinity, from first principles. We then develop a wave-optical imaging framework for a moving electromagnetic source in curved spacetime. We obtain images of an orbiting electric dipole, displaying relativistic beaming, gravitational lensing, and Einstein rings. We study polarization-dependent scattering by comparing the images produced by spinning dipoles with opposite helicities, finding a displacement that roughly decreases with the inverse of the radiation frequency, as expected for a beyond-geometric-optics effect. Our results provide a first-principles benchmark for beyond-geometric-optics descriptions of electromagnetic radiation in Kerr spacetime.
\end{abstract}
\maketitle


\section{Introduction}
Black holes (BHs) are remarkably simple astrophysical objects~\cite{Robinson:1975bv,Chrusciel:2012jk}, yet they are believed to power some of the most energetic phenomena in the Universe, including relativistic jets. Their mergers are known to release enormous amounts of energy in the form of gravitational radiation within fractions of a second. The extreme gravitational fields surrounding BHs, combined with the breakdown of classical General Relativity in their deep interiors, make them exceptional laboratories for probing the gravitational interaction and testing the limits of our current theory of gravity~\cite{Barack:2018yly}.

Support for the black-hole paradigm has grown considerably in recent years, driven by major theoretical and technological advances. Gravitational-wave observations by the LIGO--Virgo--KAGRA collaboration~\cite{LIGOScientific:2016aoc,LIGOScientific:2017vwq,LIGOScientific:2026oim} have revealed hundreds of mergers involving stellar-mass compact objects across the visible universe. In parallel, optical/infrared interferometry and very-long-baseline radio interferometry have provided unprecedented views of the center of our Galaxy and the environments of distant galactic nuclei, providing robust evidence that galaxies do harbour supermassive BHs~\cite{EHT:2019,EHT:2022SgrAshadow,Gillessen:2008qv,Ghez:1998ph}. Planned gravitational-wave detectors and upgrades to existing and next-generation telescopes in the coming years will continue to test gravity at its strongest, potentially observing sources across the full visible universe~\cite{LISA:2017pwj,Punturo_2010,Johnson:2023ynn,2022Msngr.189...17A}.

\begin{figure}[h]
\centering
\includegraphics[width=.9\linewidth]{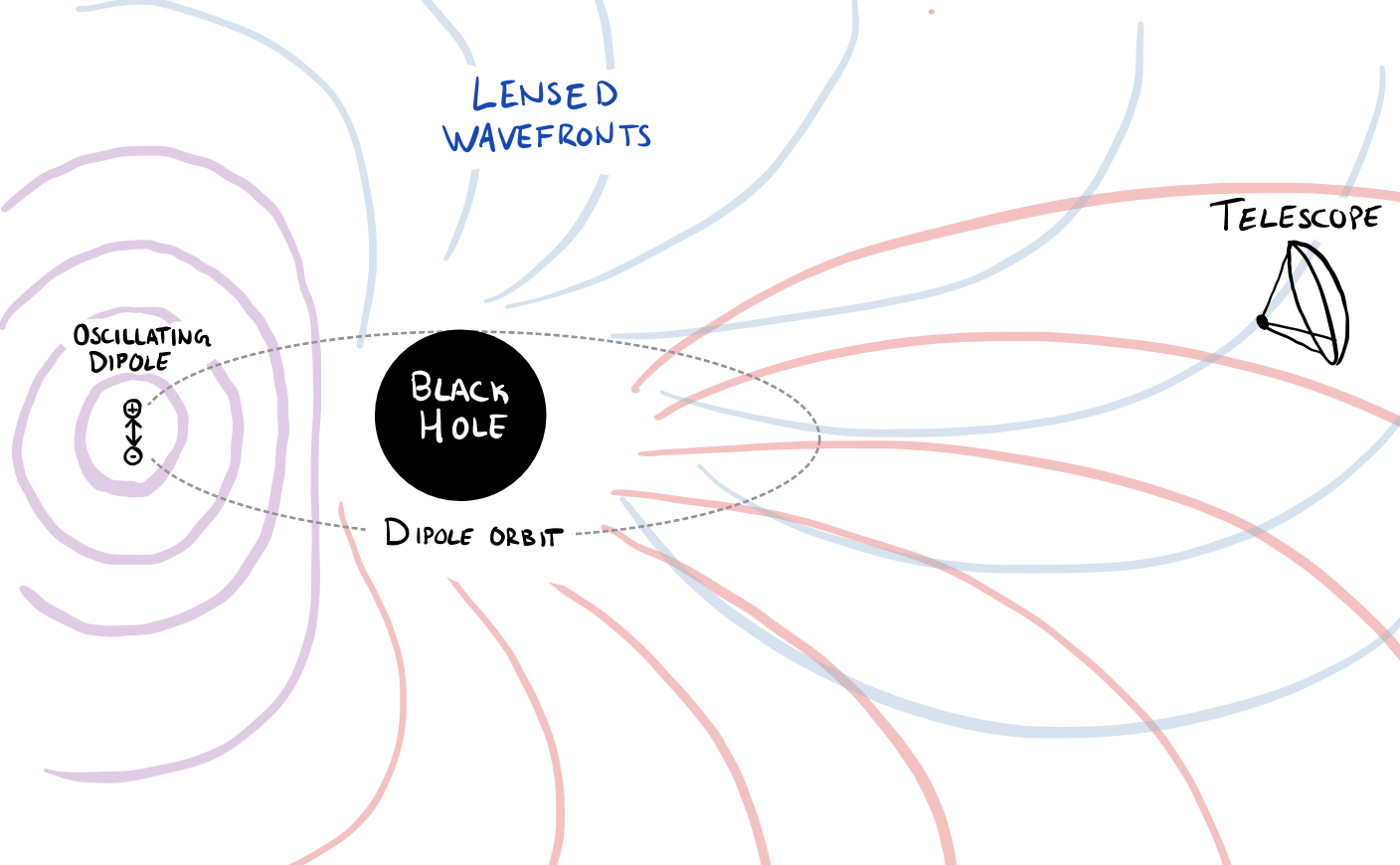}
\caption{\justifying Schematic representation of the setup studied in this paper. Not AI-generated.
}
\label{fig:orbiting_dipole}
\end{figure}

We are interested in understanding how sources of radiation appear when they are located in the strong gravitational field of a BH. The propagation and scattering of radiation from such sources in the vicinity of the BH contain valuable information about both the BH and the source itself, making the study of their perceived appearance particularly important for interpreting astrophysical observations. This is especially relevant when the source is in a close orbit around the BH, where relativistic effects can significantly alter its observed image. The challenge of BH imaging is therefore twofold: to determine how an object appears to a distant observer as it closely orbits a BH, and to understand how the resulting image encodes information about both the source and the underlying spacetime~\cite{Synge:1966okc,CunninghamBardeen1973,Luminet:1979nyg,Zhou:2024dbc}.

For supermassive BHs and optical observations, the huge disparity in scales between the radiation wavelength and the size of the compact object renders a full wave analysis impossible. Indeed, imaging in BH spacetimes has only been tackled within the geometric-optics approximation\footnote{Some noteworthy attempts do exist, focusing on simple toy models~\cite{Nambu:2015aea, Nambu:2019sqn,Willenborg:2023ixu}.}, where the electromagnetic (EM) radiation is assumed to follow null geodesics and the images are reconstructed using ray tracing ~\cite{Bozza:2002zj,Dolan:2008kf,Vincent:2011wz,Gralla:2019drh}. Such a high-frequency approximation captures many of the characteristic features of BH images, most prominently gravitational lensing. Corrections beyond geometric optics can be incorporated systematically, as higher-order terms in an expansion in powers of the inverse of the radiation frequency~\cite{Bruyere:2026gnt,Bruyere:2026xzb,Takahashi:2004mc}; in this way, one can observe effects such as polarization-dependent phenomena~\cite{Cusin:2019rmt,Oancea:2022szu,Nishida:2026bzu, Oancea:2020khc}. However, a full wave-optical treatment can capture all these in a non-perturbative manner. This is particularly important when the wavelength becomes comparable to a characteristic scale of the problem (for our purposes, the scale would be the size of the BH), where interference and diffraction can modify the image and potentially provide new probes of strong-field gravity~\cite{Ezquiaga:2025gkd, Chan:2025wgz, Li:2025lvl, Kubota:2024zkv, Motohashi:2021zyv}.

In this paper, we provide what is, to our knowledge, the first general framework for wave-optical imaging of electromagnetic (EM) radiation in a BH spacetime. We model the radiation source as a simple oscillating electric dipole, for which we provide the effective energy-momentum tensor that was previously missing from the literature. We then apply BH perturbation theory~\cite{Teukolsky:1973ha,Teukolsky:1974yv} to this source placed on an equatorial, circular orbit around a Kerr BH (see Fig.~\ref{fig:orbiting_dipole}). Our implementation, based on the framework developed in Ref.~\cite{Santos:2025ass}, allows us to compute the EM radiation entirely from first principles. For the imaging procedure, we develop a Fourier-optics approach in which the EM field is evaluated on an aperture plane far from the source and Fourier transformed to obtain the angular distribution corresponding to the image seen by a distant observer. In principle, by combining several such sources, this approach can be extended to image more general emitters in the strong-field region of a BH.

We start by setting up the dipolar radiation source in Sec.~\ref{sec:framework}, where we also solve the Teukolsky equation and benchmark our results. In Sec.~\ref{sec:imaging}, we explain how to produce images (and videos) of the dipolar (in fact, any EM) source orbiting the central BH, and show in Sec.~\ref{sec:results} that this method can capture helicity-dependent scattering. We conclude in Sec.~\ref{sec:conclusion}, where we discuss possibilities for future work.

Throughout the paper, we use geometrized units, with $G=c=1$, and adopt the mostly-plus convention, with the signature of a Lorentzian metric given by $(-1,1,1,1)$.

\section{Framework} \label{sec:framework}
%
Our goal is to model an oscillating linear electric dipole orbiting a Kerr BH and calculate the resulting EM radiation from first principles. Classically, the dipole moment of an electric dipole oscillating in the $z$ direction can be written as
\begin{equation} 
\vec{p}=p_0 \cos(\Omega_{\rm{d}} t) \hat z,\label{eq:dipEuclidian}
\end{equation}
where $p_0$ is the magnitude of the dipole moment and $\Omega_{\rm d}$ is the oscillation frequency. To study such a source in the vicinity of a rotating BH, we must generalize this description to curved spacetime and then determine the radiation emitted throughout its orbit. In this section, we lay out the necessary framework and present some benchmarking results.

\subsection{Kerr geometry and motion}
\label{subsec:kerrGeometry}

A rotating BH is described by the Kerr metric, which is given, in Boyer-Lindquist (BL) coordinates $\{t,r,\theta,\phi\}$, by~\cite{Kerr:1963ud,Bardeen:1972fi}
\begin{align}
ds^2 &= -\left(1 - \frac{2Mr}{\Sigma}\right) dt^2
- \frac{4Mar \sin^2\theta}{\Sigma} \, dt \, d\phi + \frac{\Sigma}{\Delta} \, dr^2 \notag \\
& + \Sigma \, d\theta^2
+ \left(r^2 + a^2 + \frac{2Ma^2 r \sin^2\theta}{\Sigma}\right)\sin^2\theta \, d\phi^2 \,,
\end{align}
where
\begin{equation}
\Sigma = r^2 + a^2 \cos^2\theta\,,\quad \Delta = r^2 - 2Mr + a^2\,.
\end{equation}
Here, $M$ is the BH mass and $J=aM$ is its angular momentum. We will take the worldline of the dipole to be a circular equatorial geodesic, i.e.\ a timelike geodesic with constant $r=r_0$ and $\theta=\pi/2$, given by
\begin{equation}
z(\tau) = (u^t \tau, r_0, \pi/2, u^t \Omega_0 \tau) \,, \label{eq:worldline}
\end{equation}
where
\begin{equation}
\Omega_0 = \frac{\pm\sqrt{M}}{r_0^{3/2}\pm a\sqrt{M}} \,,
\end{equation}
with the constant $u^t > 0$ fixed by the normalization condition $u^\mu u_\mu = -1$. The choice of plus or minus refers to whether the orbit is prograde or retrograde, respectively.
\subsection{Electric dipole source}
\label{subsec:source}

We want to define an effective 4-current $J^\mu$ that is conserved and  captures the dipolar nature of the source. In a generic curved spacetime, the electric dipole moment of a physical current density $J^\mu$ is, according to Dixon's formalism (formulated in Refs.~\cite{Dixon:1970zza, Dixon:1974xoz}):
\begin{equation}
p^\alpha = \int_{\Sigma(\tau,u)} x^\alpha J^\mu ~d\Sigma_\mu \,, \label{eq:dipole_moment}
\end{equation}
where the current density is supported on a worldtube around the worldline $z^\alpha(\tau)$ with unit tangent vector $u^\alpha$. The worldtube is foliated by spacelike hypersurfaces $\Sigma(\tau,u)$, each orthogonal to $u^\alpha$ at the point $z^\alpha(\tau)$. We also require the support of $J^\mu$ on each surface $\Sigma(\tau,u)$ to be compact and contained in a normal neighborhood of $z(\tau)$, where we define normal coordinates $x^\mu$.

We now want to find an effective 4-current that is supported only on the worldline $z^\alpha(\tau)$.\footnote{A more complicated approach would be to take two opposite charges in neighboring worldlines, as is done in Ref.~\cite{Lestingi:2023ovn} in the context of scalar charges.} Consider the following ansatz for that effective 4-current:
\begin{equation} \label{eq:4current}
J^\mu = 2 \int_\mathbb{R} \nabla_\alpha\left( u^{[\alpha} p^{\mu]}\delta^{(4)}(x-z(\tau)) \right) d\tau \,.
\end{equation}
It is easy to show that this 4-current is conserved, i.e.\ $\nabla_\mu {J}^\mu =0$. Moreover, if we replace the formula above in Eq.~\eqref{eq:dipole_moment}, we find it to be self-consistent, i.e.\ the right-hand side of the equation truly is equal to $p^\alpha$ (for the details, see Appendix~\ref{app:proof}). This shows that the ansatz is indeed correct.

We have to write this 4-current in a useful frame, corresponding to BL coordinates; however, it is far easier to define the dipole moment $p^\mu$ in its local frame. Thus, we need to construct frame transformations that take us from the dipole's local frame, where it is oscillating as in Eq.~\eqref{eq:dipEuclidian}, to BL coordinates, where we will compute the effective 4-current.

We begin by introducing a local frame attached to the dipole with basis vectors $\{u, e_r, e_\phi, e_z\}$, whose components are
\begin{eqnarray}
    u^\mu = u^t(\delta^\mu_t+\Omega_0 \delta^\mu_\phi)\,,\quad
    e^\mu_r = \sqrt{\frac{\Delta}{\Sigma}}\delta^\mu_r\,,\quad
    e^\mu_z = -\sqrt{\frac{1}{\Sigma}}\delta^\mu_\theta,
\end{eqnarray}
where all these quantities should be evaluated at the dipole's orbit, i.e.\ $r=r_0$ and $\theta=\pi/2$. The fourth unit vector $e_\phi$ is determined by requiring orthogonality.

Next, we consider a frame $\{u,e_1,e_2,e_3\}$ which is parallel-transported along the geodesic. If a vector $\vec e$ orthogonal to the 4-velocity is parallel-transported along a circular equatorial geodesic then, as shown in Eq.~(41) of Ref.~\cite{Rindler:1990} (see also Ref.~\cite{vandeMeent:2019cam}), it precesses with respect to the triad $\{e_r, e_\phi, e_z\}$ according to
\begin{equation}
    \frac{d\vec e}{d\tau} = \Omega_{\rm{p}} ~\vec e \times \vec e_z = \pm \sqrt{\frac{M}{r_0^3}}~\vec e \times \vec e_z \, ,
\end{equation}
where $\Omega_{\rm{p}}$ is the precession frequency (in proper time) and the plus and minus signs refer, respectively, to the prograde and retrograde orbits. The parallel-transported frame is then simply
\begin{eqnarray}
    e_1 &=&\cos(\Omega_{\rm{p}}\tau)e_r-\sin(\Omega_{\rm{p}}\tau) e_\phi \,, \\
    e_2 &=& \sin(\Omega_{\rm{p}}\tau) e_r + \cos(\Omega_{\rm{p}} \tau) e_\phi \,, \\
    e_3 &=& e_z \,.
\end{eqnarray}

Finally, we rotate this frame such that one of the spatial vectors of the new tetrad, say $E_3$, is aligned with the dipole moment $p$. Since the dipole moment is also parallel-transported, we need only to require that this condition is met initially. Then, if $\iota_{\rm{d}}$ is the inclination of $p$ relative to $e_3$, and $\lambda_{\rm d}$ is the longitude of the ascending node, the resulting frame $\{u, E_1, E_2, E_3\}$ is given by
\begin{eqnarray}
    E_1 &=& \cos(\lambda_{\rm d})~e_1 + \sin(\lambda_{\rm d}) ~e_2 \,,\\
    E_2 &=&-\cos(\iota_{\rm d})\sin(\lambda_{\rm d})~e_1 \nonumber \\
    &+& \cos(\iota_{\rm{d}})\cos(\lambda_{\rm d}) ~e_2 + \sin(\iota_{\rm{d}}) ~e_3 \,, \\
    E_3 &=& \sin(\iota_{\rm d})\sin(\lambda_{\rm d})~e_1\nonumber \\
    &-& \sin(\iota_{\rm{d}})\cos(\lambda_{\rm d}) ~e_2 + \cos(\iota_{\rm{d}}) ~e_3
\end{eqnarray}
(see Fig. S.1 in Ref.~\cite{Santos:2025ass} for a similar setup).

Finally, we can write the dipole moment in the new dipole aligned frame: 
\begin{equation}
    p^{\mu}=p_0\cos(\Omega_{\rm{d}} \tau) E_3^\mu \,,
\end{equation}
where $p_0$ is the dipole magnitude and $\Omega_{\rm d}$ is its intrinsic frequency. By performing the frame transformations above, we can now write this moment in Boyer-Lindquist coordinates.

\subsection{The resulting EM field}
%
In order to compute the EM field radiated by this source, we must solve the Teukolsky equation \cite{Teukolsky:1973ha} for the current \eqref{eq:4current}. This is a wave-type equation, and so schematically it takes the form
\begin{equation}
    \square_g \psi = 4 \pi \Sigma \mathcal{T} \,,
    \label{eq:teukSchem}
\end{equation}
where $\square_g$ is a wave operator in Kerr spacetime, $\mathcal{T}$ encodes information about the source and $\psi$ is the radiation field. In our case, and since we are interested in EM radiation, $\psi$ is the Teukolsky variable $\phi_2$, which is related to the $\hat \theta$ and $\hat \phi$ components of the electric field at future null infinity, $E_\theta$ and $E_\phi$, by
\begin{equation}
\psi = \phi_2 = \frac{E_{\theta}-iE_{\phi}}{\sqrt{2}} \,, \label{eq:phi2_Etheta_Ephi}
\end{equation}
as follows straightforwardly from Eq.~(1.1) of Ref.~\cite{Teukolsky:1973ha}.

We detail our method for solving the Teukolsky equation in Appendix~\ref{app:teuk_method}. We find that the Teukolsky variable $\phi_2$ is, asymptotically at infinity, given by
\begin{equation}
    \phi_2 \sim \sum_{\substack{\ell, m ,p,d}} {Z}^\infty _{\ell m pd} \, _{-1}S _{ \ell m \omega_{mpd}} (\theta) e^{i m \phi} \frac{e^{-i\omega_{mpd} u}}{r} \,, \label{eq:phi2_inf}
\end{equation}
where $\{u,t,r,\theta\}$ are the retarded Boyer-Lindquist coordinates, the functions $_{-1}S _{ \ell m \omega_{mpd}} (\theta)$ are the spin-weighted spheroidal harmonics \cite{Leaver1986-tf}, and the amplitudes $Z^\infty_{\ell mpd}$ are given by expression \eqref{eq:amplitudes}. This formula involves computing solutions to the homogeneous version of Eq.~\eqref{eq:teukSchem}. This must be done numerically, for which we use an implementation of the Sasaki-Nakamura method~\cite{Sasaki:1981kj,Sasaki:1981sx} in Julia programming language~\cite{Lo:2023fvv}. The frequencies excited by the system are
\begin{equation}
    \omega_{mpd}=m\Omega_0+p\Omega_{\rm{p}} + d \Omega_{\rm d} \,,
    \label{eq:frequencies}
\end{equation}
where $p \in \{0, \pm 1\}$ and $d \in \{\pm 1\}$ are the precession and dipole mode numbers, respectively. The amplitudes are determined by
\begin{equation}
    Z_{\ell m pd}^\infty = \sum_ 
    {i= 0} ^{4}\ \sum_ 
    {j = 0} ^{4-i} \mathcal{A}^{(i,j)}_{\ell m pd} \frac{d^i }{dr^i} R^{H}\left(r_0 \right) \ \frac{d^j}{d\theta^j}  \bar S \left( \pi/2\right) \, , \label{eq:amplitudes_curved}
\end{equation} 
where $\mathcal{A}$ depends only on the orbital and dipole parameters, $R^{H}\equiv R^{H}_{\ell m pd} \equiv R^{H}_{\ell m \omega_{mpd}}$ is a solution to the homogeneous radial equation~\cite{Teukolsky:1973ha}, satisfying purely
ingoing boundary conditions at the horizon, and $\bar S \equiv {_{-1}S_{\ell m \omega_{mpd}}}$.

The amplitudes $Z_{\ell mpd}$ corresponding to each frequency satisfy
\begin{equation}
    Z_{\ell(-m)(-p)(-d)} = (-1)^{\ell+p} \bar{Z}_{\ell mpd} \,.
\end{equation}
This relation follows from the symmetry of the system for the simultaneous transformations $t \to -t$ and $\phi \to -\phi$.

Finally, the energy fluxes at infinity and at the horizon are given by~\cite{Teukolsky:1974yv}, 
\begin{align}
    \dot E^\infty & = \sum_{\ell m \omega} \frac{|Z^\infty_{\ell m \omega}|^2}{2\pi} \,, \label{eq:fluxInfty}\\
    \dot E^H & = \sum_{\ell m \omega} \beta_{\ell m \omega}\frac{|Z^H_ {\ell m \omega}|^2}{2\pi} \,, \label{eq:fluxHorizon}
\end{align}
where the expression for $\beta_{\ell m p q}$ can be found in Eq.~(4.32) of Ref.~\cite{Teukolsky:1974yv}. These expressions allow us to compute waveforms and energy fluxes, which we will use in the next section to reconstruct the image of the source.

%
\subsection{Benchmarking}
\label{subsec:benchmarking}
\begin{figure}[h]
\centering
\includegraphics[width=1.0\linewidth]{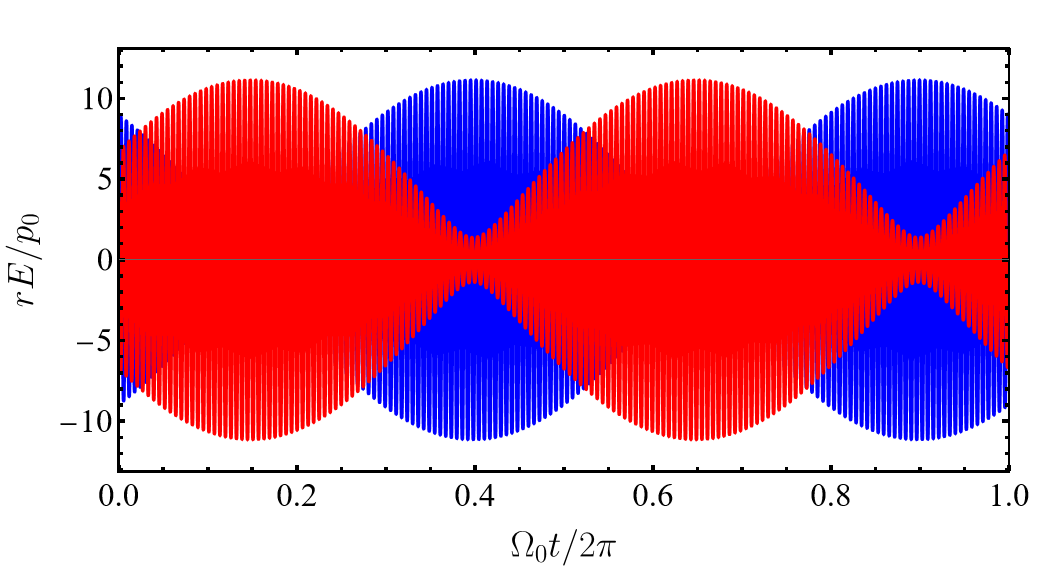}
\caption{\justifying Electric field as seen by a face-on observer, for a vertically aligned ($\iota_{\rm d}=0$) dipole of frequency $M\Omega_{\rm d}=5$ orbiting around a Kerr BH ($a=0.7$) at a distance $r_0=10M$. The different polarizations $E_\theta$ and $E_\phi$ are, respectively, in red and blue. There is an amplitude modulation with period $\pi/\Omega_{0}$; the corresponding envelopes for the two polarizations are out of phase by $\pi/2$.}
\label{fig:face_on}
\end{figure}

%
Given the absence of similar studies in the literature, it is important to thoroughly benchmark our model against known limits (such as Larmor's formula in the flat-space limit). Most of this analysis is in Appendix \ref{app:benchmarks}; here, we focus on an interesting effect that was previously misinterpreted by some of the authors in Ref.~\cite{Santos:2025ass}. The face-on signal from a dipole in orbital motion is modulated with a period half that of the orbit (Fig.~\ref{fig:face_on}). We previously attributed this to helicity-dependent scattering, but the main features are instead better explained by relativistic beaming and aberration effects.

Indeed, helicity-dependent scattering relies on spacetime curvature and is absent in Minkowski spacetime. A vertically oscillating dipole at rest in flat space emits no radiation along its axis, so a face-on observer should see no signal. Yet, an analogous calculation for a dipole in uniform circular motion in Minkowski spacetime\footnote{This means using the same methods as described in Appendix~\ref{app:teuk_method}, but now the homogeneous solutions are spin weighted spherical harmonics and spherical Bessel functions in the angular and radial sectors, respectively. The inner boundary condition becomes regularity at the origin.} produces the same modulation as above (see Fig.~\ref{fig:JoaoSuggestion} in Appendix \ref{app:faceon}), showing that it cannot be just a scattering effect.

\begin{figure}[h]
\centering
\includegraphics[width=1.0\linewidth]{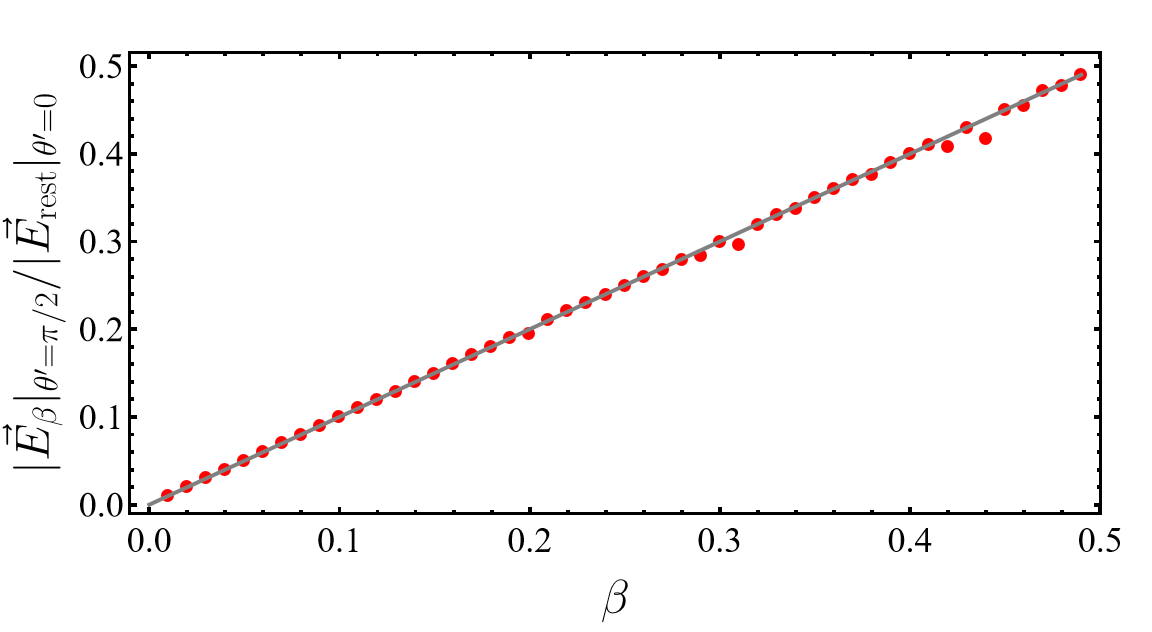}
\caption{\justifying Scaling of the face-on electric-field amplitude with the source velocity $\beta$. The gray line shows the prediction of the boosted-source model, equation~\eqref{eq:face_on_beta_scaling}, while the red points are obtained from the first-principles calculation.}
\label{fig:betaBeaming}
\end{figure}

The effect is instead largely due to aberration. While a dipole at rest emits no radiation towards the face-on observer, its radiation pattern changes when the source moves relative to the observer. The result is a face-on signal that grows with its orbital speed. For slow orbits, the maximum electric field is expected to scale linearly with $\beta$ (this prediction is derived in Appendix \ref{app:faceon}). We computed the maximum face-on electric-field strength in flat space for different orbital velocities, with the results shown in Fig.~\ref{fig:betaBeaming}. The agreement with the boosted-source prediction is remarkable, despite the approximations involved. The small systematic deviations arise from how the amplitude is measured: taking the maximum over sampled orbital periods. This can slightly underestimate the envelope maximum, which also explains why some points lie below, but none above, the prediction.

This benchmarks our results against well-understood flat-space physics, and clarifies the origin of the signal observed in previous works, such as Ref.~\cite{Santos:2025ass}.

\section{Wave optical imaging} \label{sec:imaging}

It is particularly interesting, in the case of EM radiation, to construct images of the sources themselves, as this is precisely how telescopes -- and even interferometric radio-telescope networks such as the Event Horizon Telescope -- use EM waves to detect and image astrophysical objects. In this section, we describe how such images can be obtained using Fourier optics, following the approach of Ref.~\cite{Willenborg:2023ixu}. There are two main differences in our case: the vector nature of the radiation, which requires a careful treatment of the detected field, and the motion of the source with respect to the BH, which requires us to account for multiple frequencies rather than dealing with a monochromatic source. These differences are discussed in detail in Sec.~\ref{subsec:imageConstr}. To the best of our knowledge, this is the first time both of these generalizations have been introduced, allowing for the imaging of a moving EM source in curved spacetime from first principles, rather than resorting to ray tracing.

\subsection{Fourier Optics}
\label{subsec:imageConstr}
The basic idea of Fourier optics is that, in the Fraunhofer (far-field) regime, where the observer is sufficiently far from the source that the curvature of the wavefront across the aperture can be neglected to leading order, the diffraction pattern is directly related to the two-dimensional Fourier transform of the electric field on the screen. In this regime, each transverse spatial frequency of the field corresponds to an observation direction, so that the Fourier transform determines the angular distribution of the radiation seen by the observer \cite{goodman2005introduction}. This description is the appropriate framework for describing the propagation and scattering of waves when the geometric optics approximation fails \cite{Nambu:2015aea,Willenborg:2023ixu}.
\begin{figure}[h]
    \centering
    \includegraphics[width=1\linewidth]{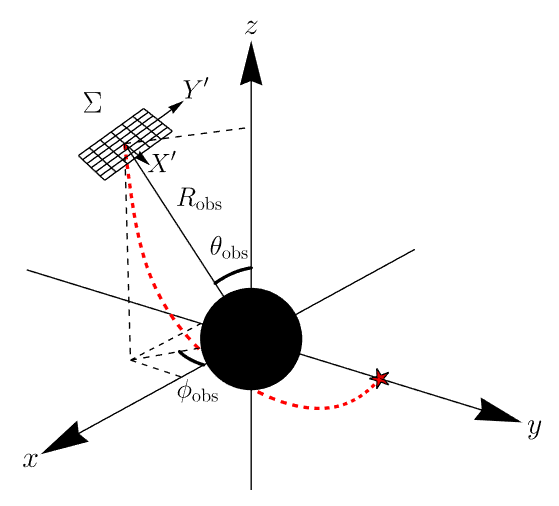}
    \caption{\justifying Diagram illustrating the imaging method. The red star represents the dipole, the black sphere represents the central BH, and the aperture, with angular coordinates $\theta_{\rm obs}$ and $\phi_{\rm obs}$, at a distance $R_{\rm obs}$ (not to scale) from the BH, is the $X'Y'$ grid labeled $\Sigma$. 
    }
    \label{fig:schematicImag}
\end{figure}

We place a small aperture at a distance $R_{\rm obs}$ from the BH, sufficiently large so that the asymptotic expression~\eqref{eq:phi2_inf} is valid, located at angular coordinates $(\theta_{\rm obs},\phi_{\rm obs})$. The aperture is the physical surface on which the EM field is sampled before the imaging procedure is applied. As a useful mental picture, one may think of light rays passing through this aperture and then being focused by an ideal optical system, producing an inverted image of the source on a separate image plane located some distance beyond the aperture. Throughout this section, however, all field calculations are performed on the aperture plane; the ``image plane'' is only a conceptual surface on which the final image is formed.

The aperture plane is oriented so that it faces the BH directly. Geometrically, it is the plane tangent to the sphere of radius $R_{\rm obs}$ at the observer's location. We introduce Cartesian coordinates $(X',Y')$ on this plane, with the $X'$-axis pointing in the direction of increasing $\phi-\phi_{\rm obs}$ and the $Y'$-axis pointing in the direction of decreasing $\theta-\theta_{\rm obs}$ (see Fig.~\ref{fig:schematicImag}).\footnote{This prescription breaks down near the poles, where the spherical basis becomes singular. There, we align the $(X',Y')$ and $(x,y)$ directions.}

In the small-angle approximation appropriate to a small, distant aperture, $X'$ and $Y'$ measure transverse distances from the line of sight. The corresponding angular coordinates are approximately
\begin{equation}
\alpha \simeq \frac{X'}{R_{\rm obs}} \,,
\qquad
\beta \simeq \frac{Y'}{R_{\rm obs}} \,.
\label{eq:smallAngleApprox}
\end{equation}

We take the aperture to be a square $\Sigma$ of side length $2D$ in this orthogonal plane,
\begin{equation}
\Sigma= \big\{ (X',Y'):\ |X'|\leq D,\ |Y'|\leq D \big\} \,.
\end{equation}
At each point $(X',Y')$ we evaluate the EM field and project it onto the local transverse basis $(\hat{\bf X},\hat{\bf Y})$ associated with the observer's image plane. This gives two polarization components,
\begin{equation}
E_X(X',Y') \,, \qquad E_Y(X',Y') \,.
\end{equation}
Then, applying the Fraunhofer approximation \cite{goodman2005introduction} to the Fresnel-Kirchoff diffraction integral \cite{sharma2006optics} gives the angular field distribution as a two-dimensional Fourier transform. Schematically,
\begin{equation}
\tilde{E}_{X,Y}\left(\alpha,\beta\right)\propto\int_\Sigma E_{X,Y}(X',Y')e^{-i\omega ( \alpha X^\prime + \beta Y')}~d\Sigma' \,,  \label{eq:Fourier_optics}
\end{equation}
where $d\Sigma' = dX' dY'$ and $\omega$ is the frequency of the radiation. Note that $\omega \alpha$ is precisely the $x$ component of the wave vector in the plane wave decomposition of the field on $\Sigma$. The observed intensity is then obtained from the Fourier-transformed field, as
\begin{equation}
I(\alpha,\beta)
\propto
|\tilde E_X|^2+|\tilde E_Y|^2 \,.
\end{equation}
This provides the basic connection between the field computed in the aperture plane and the image seen by the observer.

In what follows, instead of using angular coordinates $(\alpha,\beta)$, we use image coordinates $(X,Y)$ to describe the apparent position of the source, obtained from the former via the same small-angle approximation as in~\eqref{eq:smallAngleApprox}:
\begin{equation}
X=R_{\rm obs}\alpha \,, \qquad Y=R_{\rm obs}\beta \,. \label{eq:smallAngleApprox2}
\end{equation}
These two descriptions are equivalent, with $(X,Y)$ providing a more intuitive and useful representation of the source position. Note carefully that the coordinates $(X,Y)$ should not be confused with the coordinates $(X',Y')$ on the aperture plane introduced above. Nevertheless, we use a similar notation for the image coordinates because, under the small-angle approximation, they are related to the natural angular coordinates $(\alpha,\beta)$ by the same mapping. Thus, whenever $(X,Y)$ are used to specify the image position, they should be understood as the corresponding image coordinates obtained from $(\alpha,\beta)$ through this relation, and not as the transverse coordinates on the aperture plane.

%
\subsection{Implementation of imaging method}
In practice, the continuous Fourier transform is replaced by a discrete Fourier transform. Thus, we discretize the aperture $\Sigma$ through a $n_{\rm grid}\times n_{\rm grid}$ grid. The image quality obtained in this way comes at a high computational cost, since the field must be evaluated at every point on the grid, requiring $n_{\rm grid}^2$ independent field evaluations: for instance, a $64\times64$ grid already requires evaluating the field at $4096$ observer points, while still producing only a relatively low-resolution image. Since we are interested in producing videos of full orbits, we will have to be conservative with the number of grid points used; we found $n_{\rm grid}=16$ to be a much smaller grid number that still resolves the oscillations of the EM field.

The resolution of the final displayed image can nevertheless be increased without increasing the number of field evaluations. Following Ref.~\cite{Willenborg:2023ixu}, we zero-pad the $n_{\rm grid}\times n_{\rm grid}$ array containing the electric field by adding $n_{\rm pad}/2$ zeros on each side, obtaining an $(n_{\rm grid}+n_{\rm pad})\times(n_{\rm grid}+n_{\rm pad})$ array. Throughout this work, we will be using $n_{\rm pad}=500$. The padded array is then multiplied by a Tukey window (as in Ref.~\cite{Nambu2021-nz}), which smoothly suppresses the field towards the boundary of the computational domain before taking the Fourier transform. This reduces the sharp discontinuity between the physical grid and the artificially padded region and thereby suppresses the associated ringing and aliasing artifacts. Zero-padding does not add any new physical information: it increases the sampling density of the discrete Fourier transform, yielding a smoother representation of the same underlying angular distribution. The actual angular resolution remains determined by the extent and sampling of the physical aperture-plane data.

We should also note that the small-angle approximation \eqref{eq:smallAngleApprox} is only valid at the aperture if $D/R_{\rm obs}\ll 1$. Throughout this work, we use $D/R_{\rm obs} = 0.2$, which lies at the edge of the validity regime of the approximation: $\sin(0.2)/0.2\approx 0.993$, corresponding to a deviation of almost $1\%$. Nevertheless, we found this approximation to be sufficiently accurate.

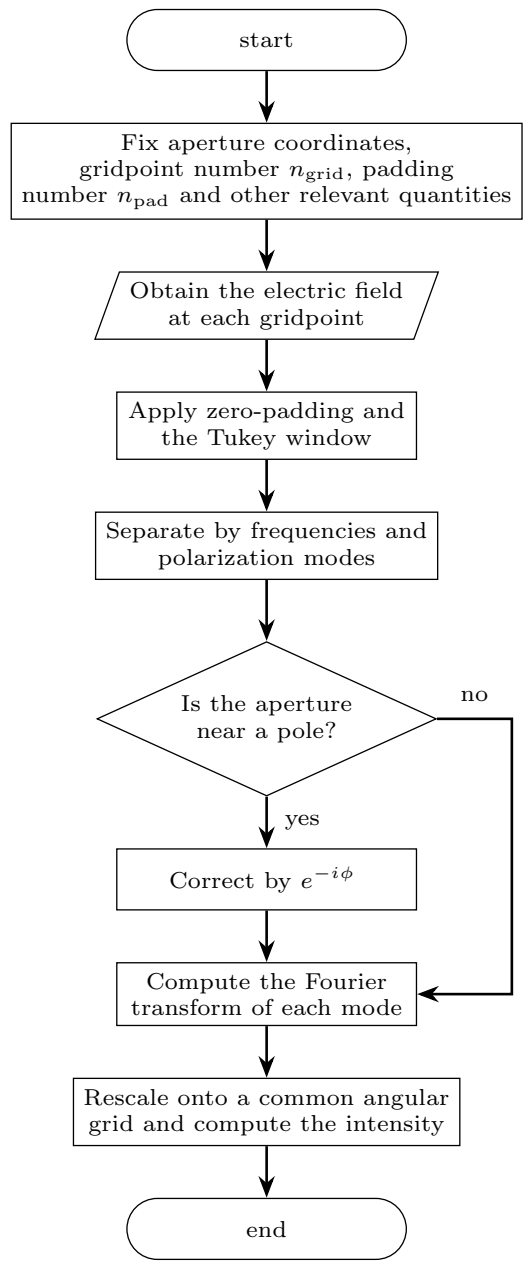
\begin{figure}[t]
    \centering
    \resizebox{0.8\columnwidth}{!}{%
\begin{tikzpicture}[
    >=Stealth,
    node distance=0.55cm,
    every node/.style={
        font=\scriptsize,
        align=center
    },
    process/.style={
        rectangle,
        draw,
        minimum width=3.2cm,
        minimum height=0.65cm
    },
    startstop/.style={
        rounded rectangle,
        draw,
        minimum width=3.2cm,
        minimum height=0.65cm
    },
    decision/.style={
        diamond,
        draw,
        aspect=2.2,
        minimum width=3.2cm,
        minimum height=0.85cm
    },
    io/.style={
        trapezium,
        trapezium left angle=70,
        trapezium right angle=110,
        draw,
        minimum width=3.2cm,
        minimum height=0.65cm
    },
    arrow/.style={
        ->,
        thick
    }
]


\node[startstop] (start) {start};

\node[process, below=0.55cm of start] (fix) {
    Fix aperture coordinates,\\
    gridpoint number $n_{\rm grid}$, padding\\
    number $n_{\rm pad}$ and other relevant quantities
};

\node[io, below=0.55cm of fix] (field) {
    Obtain the electric field\\
    at each gridpoint
};

\node[process, below=0.55cm of field] (padding) {
    Apply zero-padding and\\
    the Tukey window
};

\node[process, below=0.55cm of padding] (separate) {
    Separate by frequencies and\\
    polarization modes
};

\node[decision, below=0.65cm of separate] (near) {
    Is the aperture\\
    near a pole?
};

\node[process, below=0.55cm of near] (correct) {
    Correct by $e^{-i\phi}$
};

\node[process, below=0.55cm of correct] (fourier) {
    Compute the Fourier\\
    transform of each mode
};

\node[process, below=0.55cm of fourier] (rescale) {
    Rescale onto a common angular\\
    grid and compute the intensity
};

\node[startstop, below=0.55cm of rescale] (end) {end};


\draw[arrow] (start) -- (fix);
\draw[arrow] (fix) -- (field);
\draw[arrow] (field) -- (padding);
\draw[arrow] (padding) -- (separate);
\draw[arrow] (separate) -- (near);

\draw[arrow] (near) -- node[right=2pt] {yes} (correct);

\draw[arrow]
    (near.east)
    -- node[above=2pt] {no} ++(0.8cm,0)
    |- (fourier.east);

\draw[arrow] (correct) -- (fourier);

\draw[arrow] (fourier) -- (rescale);
\draw[arrow] (rescale) -- (end);

\end{tikzpicture}%
}
    \caption{\justifying Schematic representation of the imaging procedure.}
    \label{fig:imaging-procedure}
\end{figure}

There are, however, two important respects in which our construction differs from the scalar-field prescription of Ref.~\cite{Willenborg:2023ixu}. First, we work with a real EM field rather than a complex scalar field. This introduces two separate complications:
\begin{itemize}
    \item The first is the treatment of polarization. The electric field has two transverse components, $E_X$ and $E_Y$, which must be defined with respect to a common Cartesian basis $(\hat{\bf X},\hat{\bf Y})$. Since the field is initially computed in spherical components, whose basis vectors ($\partial_\theta, \partial_\phi$) vary across the aperture plane, a naive projection onto this Cartesian basis can introduce spurious angular effects, particularly near the poles, where the spherical basis becomes singular. The angular dependence must therefore be treated consistently when converting the field to the common Cartesian basis. In practice, when constructing an image at the poles, we rotate the complex quantity $\phi_2$ at each point in the aperture plane by $e^{-i\phi}$, ensuring the polarization basis is $(\hat {\bf X},\hat {\bf Y})$ at every point.

    \item The second complication is that the EM field in our calculation is real. If the real field were Fourier-transformed directly, its positive and negative-frequency components would both contribute. Since these components are related by complex conjugation, this would produce two copies of the same physical image, corresponding to the two sides of Fourier space. To construct a single physical image, we instead reconstruct the analytic, complex signal for each polarization component. Specifically, in the harmonic decomposition of $E_X$ and $E_Y$ (see Eqs.~\eqref{eq:phi2_Etheta_Ephi}-\eqref{eq:phi2_inf}), we retain only the positive-frequency modes, and conjugate the negative-frequency ones. This defines the complex analytic field whose Fourier transform is taken, as in Eq.~\eqref{eq:Fourier_optics}, in order to obtain the desired image. Choosing the negative-frequency modes instead (and conjugating the positive-frequency ones) would give the equivalent complex-conjugate convention; throughout this work we adopt the positive-frequency convention. 
\end{itemize}

Another noteworthy difference from the prescription in Ref.~\cite{Willenborg:2023ixu} concerns the angular coordinates of the final image. For a monochromatic plane wave with wavelength $\lambda$, the finite extent of the aperture plane determines the range of transverse wave vectors sampled by the discrete Fourier transform. For our square grid of side length $2D$ and $n_{\rm grid}$ samples per direction, the corresponding maximum viewing angle is
\begin{equation}
\theta_{\rm max}
=\frac{n_{\rm grid}\lambda}{4D}
=\frac{n_{\rm grid}\pi}{2\omega D} \,.
\end{equation}
Thus, for a fixed aperture-plane grid, the mapping between Fourier-space pixels and viewing angles depends explicitly on the frequency of the radiation.

This frequency dependence becomes important in our problem because the radiation is not described by a single observed frequency, contrary to what happens in Ref.~\cite{Willenborg:2023ixu}. The motion of the source produces Doppler shifts and, more generally, the field contains multiple harmonic contributions (see Eq.~\eqref{eq:frequencies}). A direct Fourier transform of the full time-domain field would therefore mix contributions with different wavelengths, for which the same Fourier-space pixel corresponds to different viewing angles. To avoid this, we decompose the signal into its frequency modes and perform the imaging procedure separately for each mode.

For each frequency $\omega_{mpd}$\footnote{It is worth emphasizing that, in our formalism, the Doppler effect does not need to be introduced explicitly. On the contrary, it emerges naturally through the coherent combination of the different frequency modes.}, we therefore construct the corresponding positive-frequency analytic field, evaluate its two polarization components on the aperture plane, apply the same spatial windowing and zero-padding procedure, and Fourier-transform the resulting arrays. Each frequency mode has its own effective wavelength
\begin{equation}
\lambda_{mpd}=\frac{2\pi}{\omega_{mpd}} \,,
\end{equation}
and hence its own mapping between Fourier-space coordinates and viewing angles. In particular, its maximum viewing angle is
\begin{equation}
\theta^{mpd}_{{\rm max}}=\frac{n_{\rm grid}\pi}{2 D\omega_{mpd}} \,.
\end{equation}
Therefore, the Fourier-transformed fields from different modes cannot simply be added pixel-by-pixel. Instead, before recombining the modes, we map the Fourier-space field of each frequency onto a common angular grid $(\alpha,\beta)$. In this way, a given pixel in the final image represents the same viewing direction for every frequency component. Only after performing this procedure do we sum all the Fourier-transformed fields back up, and compute the intensity.

The complete imaging procedure is represented schematically in Fig.~\ref{fig:imaging-procedure}. The first part follows the Fourier-optics construction of Ref.~\cite{Willenborg:2023ixu}, while the treatment of the two EM polarizations, the analytic-signal construction, and the frequency-dependent rescaling of the angular coordinates, required by the more general EM field considered here, are described in the second half of the diagram.

\section{Results} \label{sec:results}
%
\subsection{Imaging of an orbiting dipole}\label{subsec:imgRes}
%
%

\begin{widetext}

    \begin{figure}[h]
\centering

\begin{subfigure}{0.24\textwidth}
    \centering
    \includegraphics[width=\linewidth]{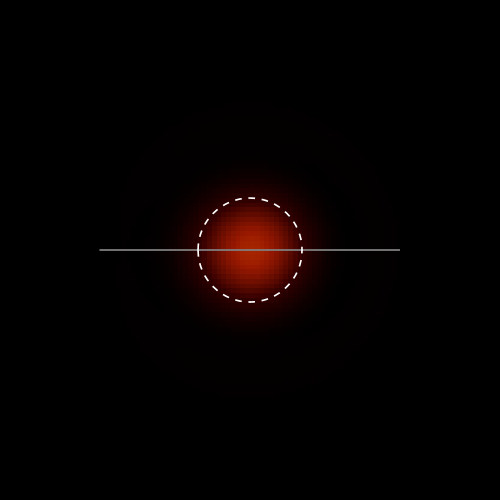}
    \caption{$\phi=0$}
\end{subfigure}
\hfill
\begin{subfigure}{0.24\textwidth}
    \centering
    \includegraphics[width=\linewidth]{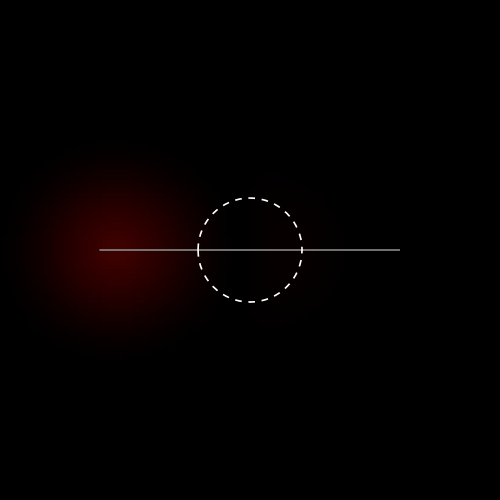}
    \caption{$\phi=\pi/2$}
\end{subfigure}
\hfill
\begin{subfigure}{0.24\textwidth}
    \centering
    \includegraphics[width=\linewidth]{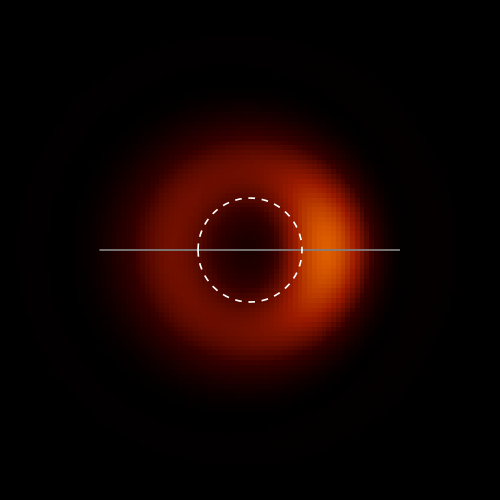}
    \caption{$\phi=\pi$}
\end{subfigure}
\hfill
\begin{subfigure}{0.24\textwidth}
    \centering
    \includegraphics[width=\linewidth]{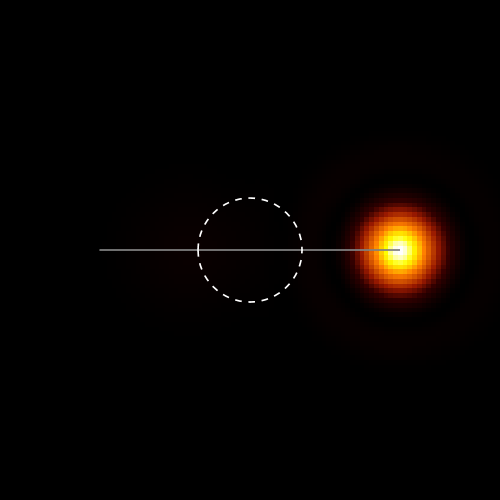}
    \caption{$\phi=3\pi/2$}
\end{subfigure}
\caption{\justifying Imaging of a dipole with frequency $M\Omega_{\rm d} = 2$ orbiting a Schwarzschild BH of mass $M$ at radius $r_0=15M$, as seen by an edge-on observer ($\theta_{\rm obs}=\pi/2$). The horizontal gray line segment represents the orbit line, while the white dashed circle corresponds to the BH shadow $r=3\sqrt{3}M$. We infer the orbital position of the dipole directly from the image. Even by eye, the signals of relativistic beaming -- compare image (b) to image (d) -- and the formation of an Einstein ring -- in figure (c) -- are clearly visible. Image (c) also shows an interesting kinematic effect: since the dipole is moving to the right at that point, relativistic beaming makes the Einstein ring appear brighter on the right side. A full video of the orbit is available at the Center of Gravity's \href{https://the-center-of-gravity.com/video-and-audio-data/}{webpage}.}
\label{fig:snapshots}
\end{figure}

\end{widetext}

We begin by showing, in Fig.~\ref{fig:snapshots}, a selection of edge-on snapshots of a dipole with frequency $M\Omega_{\rm d}=2$ as it orbits a Schwarzschild BH of mass $M$ at $r_0=15M$. It is worth noting that, due to the nature of the computation, the orbital position corresponding to each image can only be inferred from the image itself: (a) the dipole is in front of the BH moving to the left; (b) it is moving away from the observer, towards the far side of the BH; (c) it is passing behind the BH, almost diametrically opposite the observer, while moving to the right; (d) it is coming back towards the observer.

In the small angle approximation~\eqref{eq:smallAngleApprox}, the thin horizontal gray line segment in the figure extends a distance $r_0$ to each side from the origin, thus representing the orbital radius, while the white dashed circle has a radius $3\sqrt3 M$, the impact parameter of trapped photons \cite{Chandrasekhar:1985kt}, thus representing the BH shadow. It is easy to identify relativistic beaming by comparing images (b) and (d), where the source is moving away from and towards the observer, respectively. When it is moving towards the observer, the image becomes brighter, as opposed to when it is moving away. It is also clear that, as the dipole gets directly behind the BH, an Einstein ring is formed, surrounding the BH and forming its shadow. In all plots, the color maps to intensity, with black representing zero, white denoting the maximum, and intermediate values rendered through an orange gradient.

The apparent angular size of the dipole is set primarily by the diffraction limit of the lens, $\Delta\alpha = \Delta\beta\sim\lambda_{mpd}/D$, which implies that radiation modes with a longer wavelength will appear more ``smeared out'', while larger frequencies will appear more localized. This is evident, for example, in snapshots (b) and (d), where the radiation is Doppler red- and blueshifted, respectively.

\begin{figure}[h]
    \centering
    \includegraphics[width=0.9\linewidth]{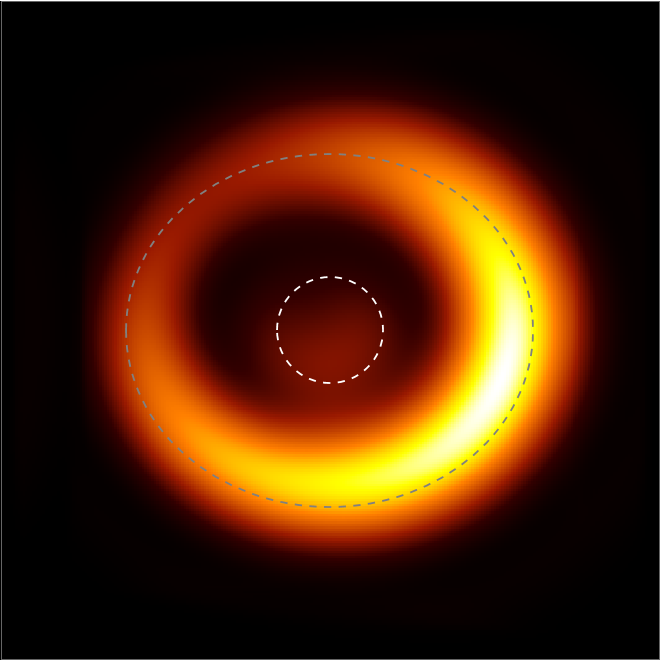}
    \caption{\justifying Integrated image corresponding to the $r_0=20M$ circular orbit of a $M \Omega_{\rm d} =2$ electric dipole around a Schwarzschild BH of mass $M$, as seen by an observer at $\theta_{\rm obs}=\pi/6$. The gray dashed line shows what the orbit would look like in flat spacetime, while the white dashed line represents the BH shadow. Relativistic beaming and gravitational lensing are evident, respectively, by the asymmetric brightness between the rightmost and leftmost portions of the orbit, and a secondary image forming near the BH shadow.}
    \label{fig:exposition}
\end{figure}

Besides individual snapshots, we can also capture what an entire orbit looks like by integrating intensities over an entire orbit. One such image is depicted in Fig.~\ref{fig:exposition}, which corresponds to a dipole of frequency $M\Omega_{\rm d}=2$ at radius $r_0=20M$, orbiting a Schwarzchild BH, and an observer looking from an angle $\theta_{\rm obs}=\pi/6$. It is worth noting the resemblance between this image and those recorded by the Event Horizon Telescope of the M87 and SgrA* BHs~\cite{EHT:2019, EHT:2022SgrAshadow}.

The gray dashed line in Fig.~\ref{fig:exposition} now corresponds to what the orbit would look like in flat spacetime, while the white dashed line still represents the BH shadow. We identify, again, the signs of relativistic beaming (the rightmost part of the orbit is much more intense than the lefttmost part) and gravitational lensing (there are signs near the BH shadow that a secondary image is formed when the dipole is furthest from the observer).

\subsection{The optical Magnus/gravitational spin Hall effect}
It is a well-known result that the propagation of EM (and gravitational) waves in curved spacetime depends on their polarization~\cite{Costa:2018gva,Nishida:2026bzu}. We provide evidence that our computation captures this physical phenomenon, further demonstrating the robustness of our calculation. We focus on a Schwarzschild background, but our methods apply more generally to Kerr spacetimes.

\begin{figure}[h]

    \begin{subfigure}{0.24\textwidth}
    \centering
    \includegraphics[width=\linewidth]{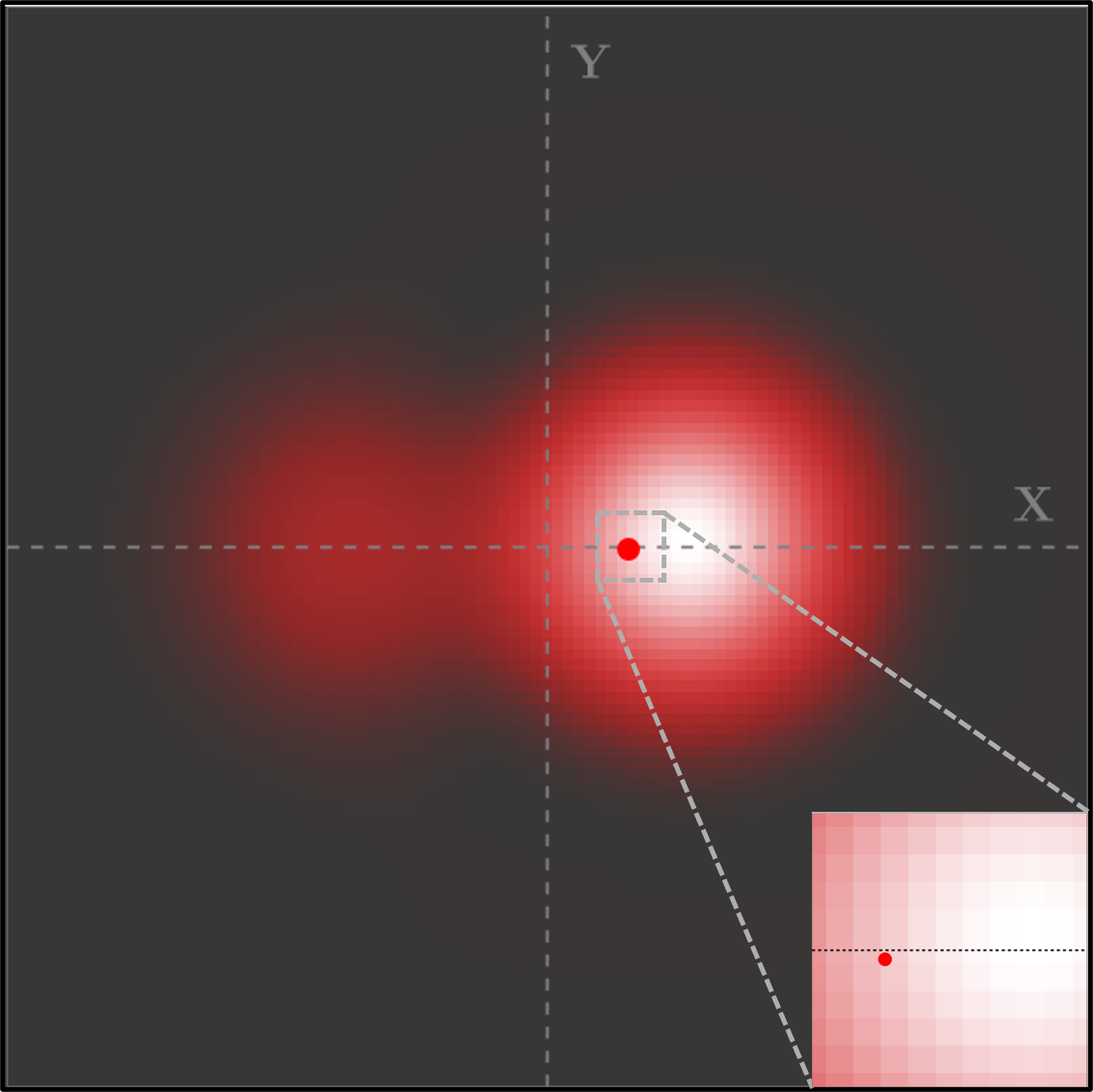}
    \caption{Negative helicity}
\end{subfigure}
\hspace{-1.5mm}
    \begin{subfigure}{0.24\textwidth}
    \centering
    \includegraphics[width=\linewidth]{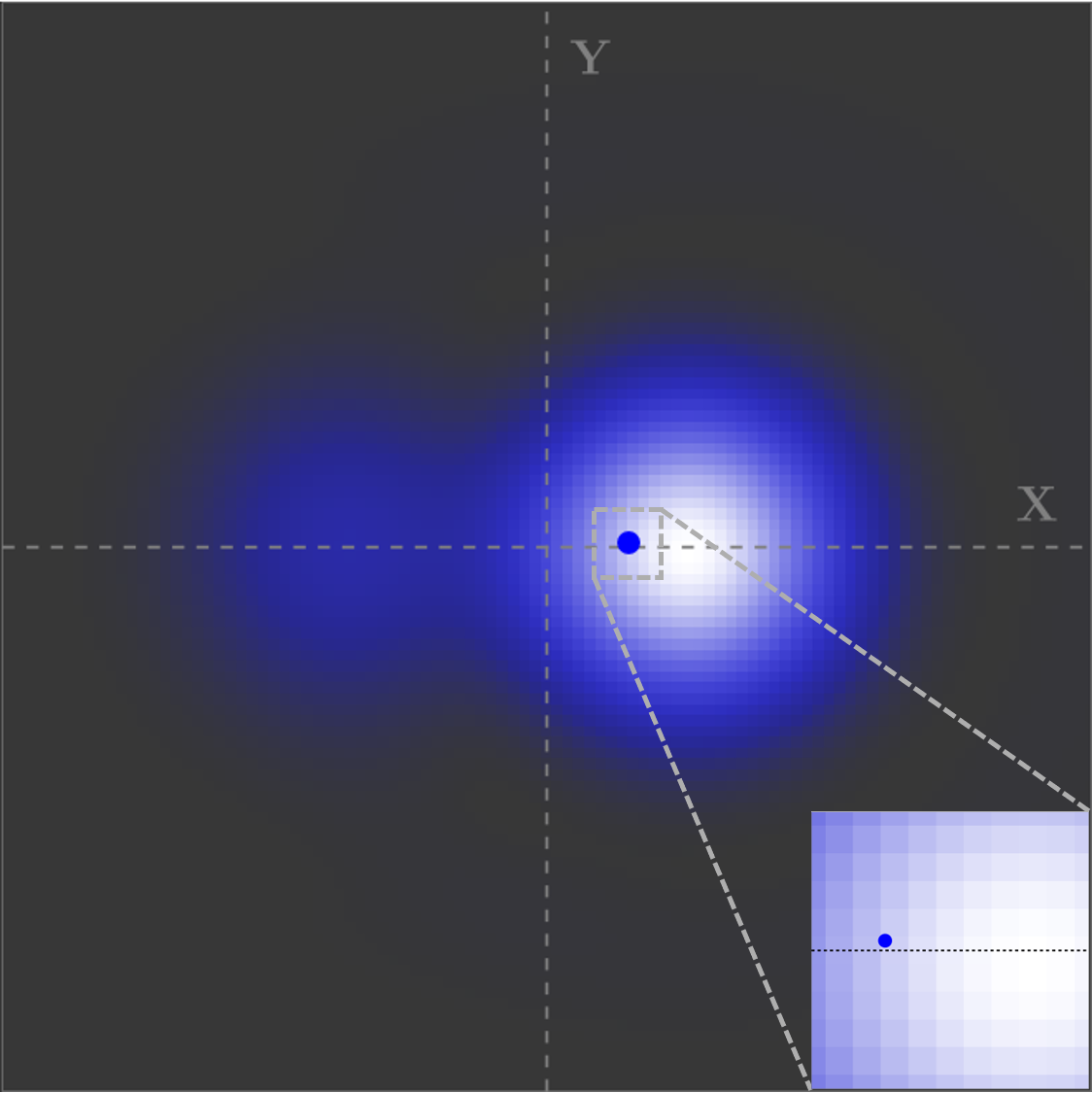}
    \caption{Positive helicity}
\end{subfigure}

    \begin{subfigure}{0.45\textwidth}
    \centering
    \includegraphics[width=\linewidth]{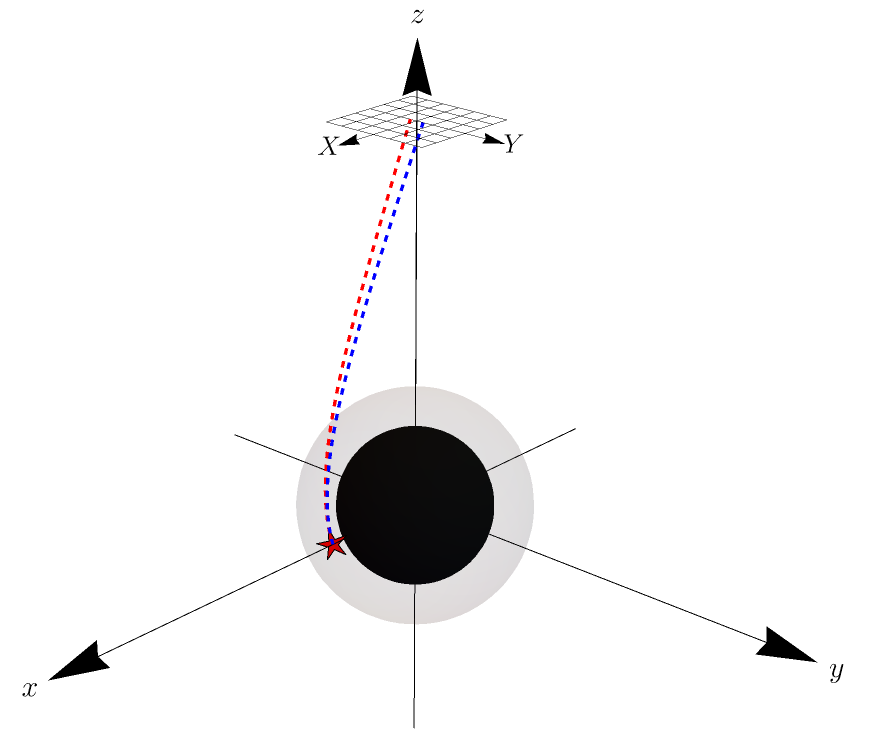}
    \caption{Schematic for the setup}
\end{subfigure}

\caption{\justifying Helicity-dependent scattering setup and images. (c) A Schwarzschild BH (black sphere) and its photon sphere (gray sphere) with a spinning dipole (red star) emitting positive (blue) or negative (red) helicity radiation along $z$. The rays do not follow the same null geodesic, but take instead distinct paths symmetric about the $xz$ plane. (a)–(b) Computed images for negative and positive helicities ($M\Omega_{\rm d}=2.6$). The zoomed regions highlight the respective centroid shifts at $(4.59M,-0.21M)$ and $(4.59M,0.21M)$, respectively, showcasing the helicity-dependent propagation.}
    \label{fig:polDepSketch}
\end{figure}

We consider a spinning dipole at rest ($\Omega_0=0$) on the photon sphere ($r_0=3M$), where this effect is most prominent. To construct such a configuration, we superimpose the radiation emitted by two dipoles oscillating linearly in different directions: one in the $\hat{r}$-direction (that is, $\iota_{\rm d} = \pi/2$ and $\lambda_{\rm d} = \pi/2$), and the other in the perpendicular $\hat{\bf \phi}$-direction (that is, $\iota_{\rm d} = \pi/2$ and $\lambda_{\rm d} = 0$). The second dipole is out of phase with the first one by $\pi/2$, so that, if one of the dipolar moments is proportional to $\cos(\Omega_{\rm d} t)$, the other one will be proportional to $\sin(\Omega_{\rm d} t)$. In this way, we effectively obtain the radiation emitted by a rotating dipole moment at rest on the photon sphere (see Fig.~\ref{fig:polDepSketch}).


It is a textbook result that the radiation emitted in the direction normal to the rotation plane of the dipole vector in flat space is (in the far-field region) circularly polarized~\cite{Jackson:1998nia}. Thus, in our setup, a face-on observer will detect nearly circularly polarized radiation, with helicity given by the direction of rotation of the dipole. 

In curved spacetime, the null rays acquire a correction to geodesic motion that depends on the helicity of the wave (see, for example, Ref.~\cite{Nishida:2026bzu}). The explicit expressions for the trajectory of the photon and the evolution of the wave vector are
\begin{align}
    \frac{d k_i}{dt} &= -k\frac{x^i}{r}\partial_r v \,, \label{eq:polDep1} \\
    \frac{d x_i}{dt} &= v\frac{k_i}{k}
    - \lambda\varepsilon^{ijl}\frac{x^j}{r}\frac{k_l}{k^2}\partial_rv \,, \label{eq:polDep2}
\end{align}
where ${\bf k}$ is the wave vector, $\lambda=\pm 1$ whether the photon is right-handed (positive helicity) or left-handed (negative helicity), and $v=(1-M/2r)/(1+M/2r)^3$. The last term in Eq.~\eqref{eq:polDep2} is responsible for this helicity-dependent correction; it goes to $0$ as $1/k$ at large frequencies.

Eqs.~\eqref{eq:polDep1}-\eqref{eq:polDep2} imply that the image of a rotating dipole detected by a face-on observer will be shifted due to helicity-dependent scattering. More importantly, the shift is in opposite directions for positive and negative helicity waves. We expect this effect to disappear at higher dipole frequencies, since $k \sim \Omega_{\rm d}$.

In order to estimate this effect, we compare two images produced by a dipole with the same frequency, at rest at fixed $r_0$, but with opposite helicities. For each image, we compute its centroid,
\begin{equation}
    (\alpha_c,\beta_c)=\frac{1}{L}\iint_\Sigma (\alpha, \beta)~I(\alpha, \beta) ~d\alpha \, d\beta \,,
    \label{eq:centroid}
\end{equation}
where the integral is performed on the field-of-view $\Sigma$ of the observer and $L$ is the apparent surface brightness of the image,
\begin{equation}
    L=\iint_\Sigma I(\alpha,\beta) ~d\alpha \, d\beta \,.
\end{equation}
We can then convert the angular coordinates of the centroid into apparent distances using Eq.~\eqref{eq:smallAngleApprox2}. If the images produced with different helicities are indeed different, as Eq.~\eqref{eq:polDep2} predicts, we will see a shift of the form $(0,\delta Y)$ in the position of the centroid for both images.

Fig.~\ref{fig:polDepSketch} showcases an example of this procedure. In both images, the dipole is at rest on the photon sphere $r=3M$, and the dipolar moment is spinning with frequency $M \Omega_{\rm d}=2.6$, clockwise in image (a) and counter-clockwise in image (b). This produces, respectively, negative and positive helicity radiation for a face-on observer, which propagate differently, as suggested by Eq.~\eqref{eq:polDep2}. It is difficult to see the difference with the naked eye, but, looking at the centroid of each image (represented by a red/blue dot), it is clear that there is a difference between the two pictures: indeed, while in image (a) the centroid is located roughly at $(4.59M,0.21M)$, in image (b) it is at $(4.59M,-0.21M)$. Thus, in this case, the displacement is $\delta Y \approx 0.42M$.

\begin{figure}[h]
    \centering
    \includegraphics[width=1.\linewidth]{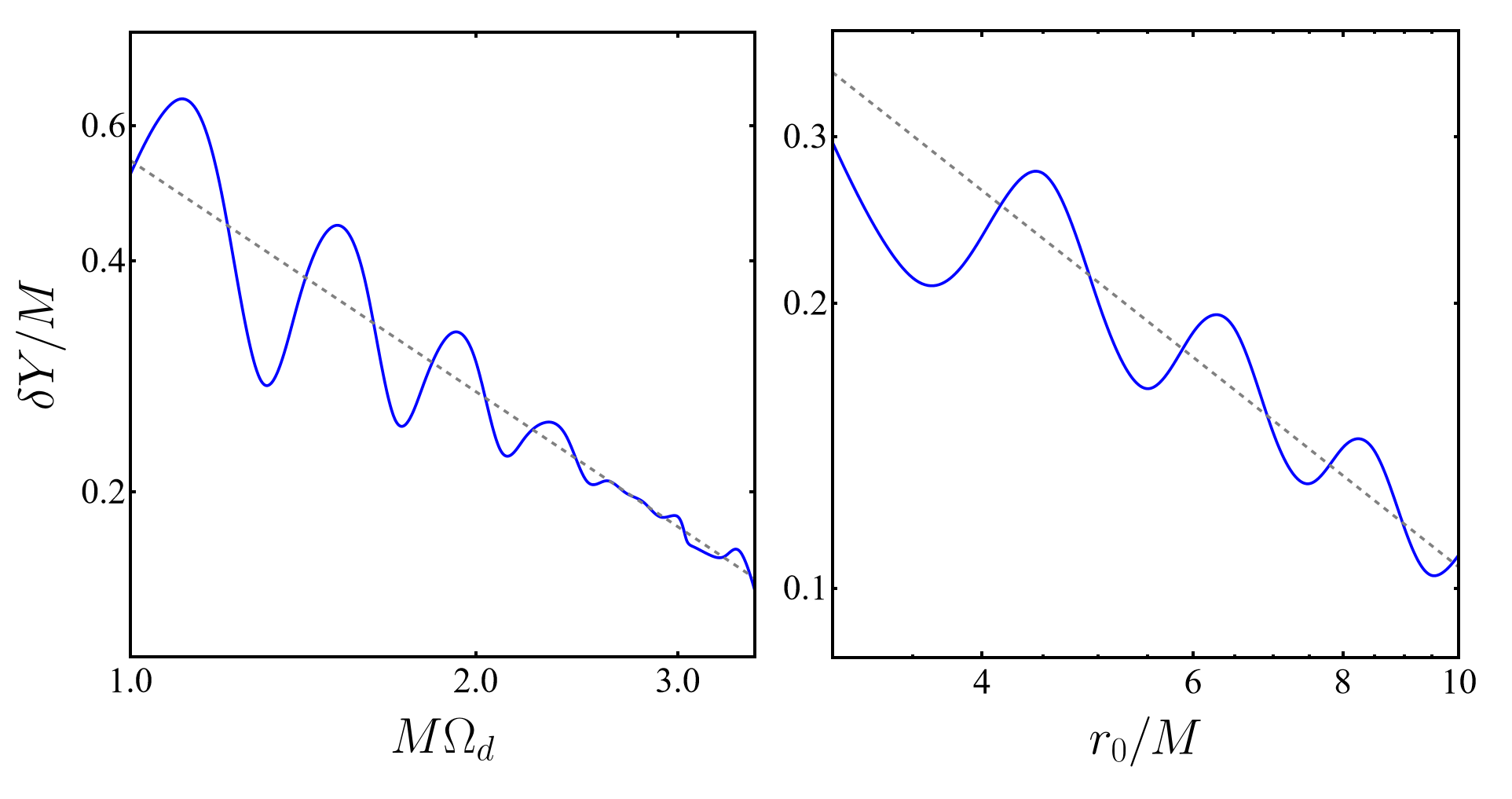}
    \caption{\justifying Log-log plots of the displacement between the positions of the centroids of the images of the clockwise and counterclockwise rotating dipoles, $\delta Y$, as a function of dipole rotation frequency, $\Omega_{\rm d}$ (left), and distance to the BH, $r_0$ (right). The difference tends to $0$ as $1/\Omega_{\rm d}$ and $1/r_0$, respectively; the oscillating nature of the decay is not yet well understood.}
    \label{fig:polDepScatt}
\end{figure}

Repeating this procedure for different frequencies leads to the leftmost plot in Fig.~\ref{fig:polDepScatt}, which depicts the displacement $\delta Y$ between the centroids of the two images as a function of the dipole frequency $\Omega_{\rm d}$. Our results are consistent with the prediction implied in Eq.~\eqref{eq:polDep2}, with the displacement approaching zero as $1/\Omega_{\rm d}$ (dashed line). We also investigated a different setup by keeping the frequency constant $M\Omega_{\rm d}=2$ and moving the dipole away from $r_0=3M$. In this case, one can show\footnote{Using the approximations $(x(t),y(t),z(t)) \sim (r_0,0,t)$, ${\bf k} \sim (0,0,k)$ and $\partial_rv \sim 2M/r^2$ yields a difference term proportional to $$\int_0^{+\infty} \frac{r_0}{\left(r_0^2 + t^2\right)^{\frac32}} \, dt = \frac1{r_0}.$$} that the displacement should also go to $0$ as $r_0$ increases, approximately as $1/r_0$, which is exactly what we show in the rightmost plot of Fig.~\ref{fig:polDepScatt}. Interestingly, however, the displacement appears to oscillate around this prediction rather than approach it monotonically, in both setups. This behavior could have several origins, including BH resonances~\cite{Santos:2026lzq}, and warrants further investigation.

\section{Conclusions} \label{sec:conclusion}
We have developed the first general framework for computing and imaging the electromagnetic (EM) radiation emitted by a source orbiting a Kerr black hole, entirely from first principles. A key ingredient is the derivation of an effective $4$-current for a pointlike, linearly oscillating electric dipole in curved spacetime, previously missing in the literature. Using this current as the source, we solve the Teukolsky equation for a dipole moving in a circular equatorial orbit around a Kerr black hole, obtaining the full frequency-domain EM field. We further develop a Fourier-optics imaging framework capable of handling moving, polarized EM sources in curved spacetime, vastly extending existing Fourier-optics approaches, which have so far been restricted to static scalar sources.

The resulting radiation exhibits the expected signatures of strong-field propagation: relativistic beaming, Doppler shifts, and gravitational lensing, including the formation of Einstein rings and secondary images. Importantly, we found that a face-on observer sees a modulated signal, and our analysis shows that it is largely a kinematic consequence of relativistic beaming and aberration, rather than helicity-dependent scattering, as previously stated in Ref.\cite{Santos:2025ass}. Our calculation also provides a direct test of helicity-dependent scattering: by comparing radiation from clockwise- and counterclockwise-rotating dipoles, at rest with respect to the BH, we find a displacement between the corresponding images. For constant distance $r_0$, and varying the dipole frequency $\Omega_{\rm d}$, we find that this displacement decreases as $1/\Omega_{\rm d}$, while for constant dipole frequency and variable distance we find that it decreases as $1/r_0$. These results are consistent with the predictions of helicity-dependent corrections to geometric-optics propagation (see Ref.~\cite{Nishida:2026bzu}). 

We note, however, that our dipole sources are not likely to represent any astrophysical object. Even when superimposing multiple oscillating dipoles, the frequencies that we can numerically solve for remain far below those relevant for realistic astrophysical sources: in the best case scenario, the source's frequency can only go up to $M\Omega_{\rm d} \sim 10$, which for a stellar mass BH ($M\approx 3M_\odot$ being, again, the best case scenario) corresponds to $\Omega_{\rm d} \sim 10^5 ~\rm Hz$, that is, a low-frequency radio wave, suitable for AM broadcasting. Nevertheless, because our results are computed from first principles, they provide a valuable benchmark for beyond-geometric-optics approaches. In particular, they can be used to quantify the errors resulting from truncating the high-frequency expansion at a finite order.

Natural next steps for this work are extending the calculation to more realistic sources of radiation, such as an isotropically emitting hotspot, in order to extrapolate centroid tracks and brightness curves, and compare them to classical geometric-optics results such as Ref.~\cite{Hamaus:2008yw}. It is also worth exploring the implications for rapidly spinning BHs, where polarization rotation effects can lead to more exciting phenomenology \cite{Frolov:2024qow, Frolov:2025bva}.

\begin{acknowledgments}

We thank Enrico Barausse, Paolo Pani and Luka Vujeva for fruitful discussions. We especially thank Rico Lo for helping with his Julia code, and João L.~Rosa for running GYOTO to benchmark our results against geometric optics. 
We also acknowledge the use of the Black Hole Perturbation Toolkit \cite{BHPToolkit} and xAct \cite{xAct, Martin-Garcia:2007bqa}.
The Center of Gravity is a Center of Excellence funded by the Danish National Research Foundation under grant No. DNRF184.
We acknowledge support by VILLUM Foundation (grant no. VIL37766).
V.C.\ is a Villum Investigator and a DNRF Chair.  
V.C. acknowledges financial support provided under the European Union’s H2020 ERC Advanced Grant “Black holes: gravitational engines of discovery” grant agreement no. Gravitas–101052587, and also by the H2020-MSCA-2022-SE
project EinsteinWaves, grant agreement no. 101131233.
Views and opinions expressed are however those of the author only and do not necessarily reflect those of the European Union or the European Research Council. Neither the European Union nor the granting authority can be held responsible for them.
This project has received funding from the European Union's Horizon 2020 research and innovation programme under the Marie Sklodowska-Curie grant agreement No 101007855 and No 101131233.
This work is supported by Simons Foundation International \cite{sfi} and the Simons Foundation \cite{sf} through Simons Foundation grant SFI-MPS-BH-00012593-11.
J.N. was funded by FCT/Portugal and the Recovery and Resilience Plan (PRR) through projects UID/04459/2025 and UID/PRR/04459/2025.
\end{acknowledgments}

\clearpage

\appendix

\begin{widetext}
%
\section{Proof for the $4$-current of a dipole} \label{app:proof}
%
In this section, we show that the $4$-current given by Eq.~\eqref{eq:4current} is consistent with Eq.~\eqref{eq:dipole_moment}:

    \begin{align*}
p^\beta
&= \int_{\Sigma_t} x^\beta
\left(
\int_\mathbb{R}
\nabla_\alpha
\left(
(u^\alpha  p^\mu - u^\mu  p^\alpha)
\delta^{(4)}(x-x(\tau))
\right)
\,d\tau
\right)
\,d\Sigma_\mu
= \int_{\Sigma_t} x^\beta \nabla_\alpha
\int_\mathbb{R}
(u^\alpha  p^\mu - u^\mu  p^\alpha)
\delta^{(4)}(x-x(\tau))
\,d\tau
\,d\Sigma_\mu
\\
&= \int_{\Sigma_t} x^\beta \nabla_\alpha
\left[
\left(
(u^\alpha  p^\mu - u^\mu  p^\alpha)
\delta^{(3)}(\vec{x}-\vec{x}(t))
\right)
\frac{d\tau}{dt}
\right]
u_\mu\,d^3x
=-\int_{\Sigma_t}
\nabla_\alpha(x^\beta u_\mu)
\left(
(u^\alpha  p^\mu - u^\mu  p^\alpha)
\delta^{(3)}(\vec{x}-\vec{x}(t))
\right)
\frac{d\tau}{dt}
\,d^3x
\\
&= \int_{\Sigma_t}
\delta_\alpha^\beta u_\mu
(u^\alpha  p^\mu - u^\mu  p^\alpha)
\delta^{(3)}(\vec{x}-\vec{x}(t))
\frac{d\tau}{dt}
\,d^3x
= \int_{\Sigma_t}
u_\mu
(u^\beta  p^\mu - u^\mu  p^\beta)
\delta^{(3)}(\vec{x}-\vec{x}(t))
\frac{d\tau}{dt}
\,d^3x
\\
&= \int_{\Sigma_t}
u_\mu u^\mu  p^\beta
\delta^{(3)}(\vec{x}-\vec{x}(t))
\frac{d\tau}{dt}
\,d^3x
= \int_{\Sigma_t}
 p^\beta
\delta^{(3)}(\vec{x}-\vec{x}(t))
\frac{d\tau}{dt}
\,d^3x
=  p^\beta \,,
\end{align*}
as we wished to show.
\end{widetext}

%
\section{Method for solving Teukolsky's equation} \label{app:teuk_method}
%
In this section, we show how we solve Teukolsky's equation with the effective $4$-current given in Eq.~\eqref{eq:4current} as a source. We use the Kinnersley tetrad~\cite{Kinnersley:1969zza} $(l^\mu, n^\mu, m^\mu, \bar{m}^\mu)$, where $l_\mu n^\mu = -m_\mu \bar{m}^\mu = -1$, and all other inner products vanish. Explicitly:

\begin{equation}
    \begin{cases}
        l^\mu = \left(\frac{r^2+a^2}{2}, 1, 0 \frac{a}{\Delta} \right) \\
        n^\mu = \frac{1}{2\Sigma} \left(r^2+a^2,-\Delta, 0, a \right) \\
        m^\mu = \frac{1}{\sqrt2 (r+ia\cos\theta)} \left( ia \sin\theta, 0, 1, \frac{i}{\sin \theta}\right)
    \end{cases} \,.
\end{equation}

To obtain the radiation at infinity, we compute the master variable $\phi_2$, with spin weight $-1$. Decomposing the field and source into Fourier-harmonic components, we have
\begin{align}
    \rho^{-2} \phi_2& = \int_{\mathbb{R}} d\omega \sum_{\ell,m} R_{\ell m\omega}(r){}_{-1}S_{\ell m\omega}(\theta) ~e^{im\phi-i\omega t} \,, \\
    4\pi\Sigma\mathcal{T} &= \int_{\mathbb{R}} d\omega \sum_{\ell,m} \tilde T_{\ell m\omega}(r){}_{-1}S_{\ell m\omega}(\theta) ~e^{im\phi-i\omega t} \,,
    \label{eq:fourierComps}
\end{align}
where $_{-1}S_{\ell m\omega}(\theta)$ are spin $-1$ spheroidal harmonics and $\rho=-(r-ia\cos\theta)^{-1}$. Then, omitting indices, the radial equation reads:

\begin{equation} \label{radialTeukolsky}
   \left(\Delta \frac{d^2} {dr^2} + V(r)\right) R = - \tilde T \,,
\end{equation}
where
\begin{equation}
    V(r) = \frac{K^2+2i(r-M)K}{\Delta} - 4i\omega r \,,
\end{equation}    
\begin{equation}
    K=(r^2+a^2)\omega-am 
\end{equation}
and $\lambda = A+a^2\omega^2-2am\omega$. We start by solving the homogeneous version of \eqref{radialTeukolsky},
\begin{equation}
    \left(\Delta \frac{d^2} {dr^2} + V(r)\right) R = 0 \,.
\end{equation}
This equation is of second order, and thus admits two linearly independent solutions. In particular, there solution $R^\infty $ and $R^H$ which correspond to purely outgoing radiation at infinity and purely ingoing radiation on the horizon, respectively. Their behaviors as $r \to \infty, r \to r_+ \equiv M + \sqrt{M^2 -a^2}$ are, respectively
\begin{equation}
    \begin{cases}
        R^H \sim A_{\text{in}} r^{-1} e^{-i\omega r_*} + A_{\text{out}} ~r e^{i\omega r_*} \\
        R^\infty \sim r e^{i\omega r_*}
    \end{cases} (r\to \infty) \label{eq:asymptoticsInfinity}
\end{equation}
and
\begin{equation}
    \begin{cases}
        R^\infty \sim B_{\text{in}} \Delta e^{-ikr_*} + B_{\text{out}} e^{ikr_*} \\
        R^H \sim \Delta e^{-ikr_*}
    \end{cases} (r \to r_+) \,,
    \label{eq:asymptoticsHorizon}
\end{equation}
with $r_*$ the usual tortoise coordinate and $k = \omega - m \Omega_H$, where $\Omega_H = \frac{a}{2Mr_+}$ is the horizon angular velocity. For these solutions, the Wronskian $W$ is constant 
\begin{equation}
    W = R^\infty (R^H)' - R^H (R^\infty)' = -2i\omega A_{\text{in}} \,.
    \label{eq:wronskian}
\end{equation}
Then, the solution to the non-homogeneous equation in the asymptotic regions at infinity and at the horizon is
\begin{equation}
    R_{\ell m\omega} \sim \begin{cases}
        Z^\infty_{\ell m\omega} r e^{i\omega r_*} \quad (r \to \infty) \\
        Z^H_{\ell m\omega} \Delta e^{-ikr_*} \quad (r\to  r_+)
    \end{cases},
\end{equation}
where, ommiting indices,
\begin{equation}
    Z^{\infty,H} = \frac{1}{W} \int_{r_+}^\infty \frac{1}{\Delta} R^{H,\infty}(r') \tilde T(r') dr' \,.
\end{equation}

Turning our attention now to the source term in the Teukolsky equation, it is given by $\mathcal{T} = \rho^{-2}J_2$, where the expression for $J_2$ can be found in Eq.~(3.8) of Ref.~\cite{Teukolsky:1973ha}. That expression can be written in compact form as
\begin{equation}
    J_2 = D_{\bar{m}}J_{\bar{m}} + D_n J_n \,, 
\end{equation}
where $D_{\bullet}$ are first order differential operators and $J_\bullet$ are projections of the 4-current (see Eq.~\eqref{eq:4current}) onto the Kinnersley tetrad. We can then write 
\begin{equation}
    \tilde T_{\ell m\omega} = 2 \int_\mathbb{R} e^{i\omega t'} \int_{\mathbb{S}^2} {}_{-1}\bar{S}_{\ell m\omega} e^{-im\phi'} \frac{\Sigma J_2}{\rho^2} ~d\Omega' dt'\,,
\end{equation}
which gives
\begin{equation}
    Z^{\infty,H} = \int_\mathcal{M} \mathcal{F}^{\infty, H} e^{i(\omega t' - m\phi')} J_2(x') ~d^4x' \,,
\end{equation}
where $\mathcal{F}^{\infty, H} = \frac{2 \Sigma \sin\theta'}{W\Delta \rho^2} R^{H, \infty}(r') \bar{S}(\theta')$, and the integration region is $\mathcal{M} = \mathbb{R} \times (r_+,+\infty) \times \mathbb{S}^2$.

Our effective dipole will be moving on the wordline of an equatorial geodesic of the Kerr BH~\eqref{eq:worldline}, and the $4$-current $J_\mu$ is given in \eqref{eq:4current}. Then it is possible to show that
\begin{align}
    Z^{\infty,H} &= \int_\mathcal{M} \mathcal{F}^{\infty, H} e^{i(\omega t' - m\phi')} J_2(x') ~d^4x' \label{eq:amplitudeBeforeParts} \\
    &= \int_\mathcal{M} \left( D_{\bar{m}} J_{\bar{m}} + D_n J_n\right)\mathcal{F}^{\infty, H} e^{i(\omega t' - m\phi')} ~d^4x' \nonumber \\
    &= \int_\mathcal{M} \left(J_{\bar{m}} D^\dagger_{\bar{m}} + J_n D^\dagger_n\right) \left(\mathcal{F}^{\infty, H} e^{i(\omega t' - m\phi')} \right) ~d^4x' \,, \nonumber \label{eq:amplitudeBeforeParts}
\end{align}
where $D^\dagger_{\bar{m}}, D^\dagger_n$ are the formal adjoints of $D^{\bar{m}}$ and $D_n$. The final step is to simplify $J_n$ and $J_{\bar{m}}$. This is important because, as equation~\eqref{eq:Jn} shows, these terms have covariant derivatives acting on Dirac delta functions. We will perform the calculation only for $J_n$, as it is quite similar for the other term. First, notice that
\begin{equation}
    J_n = n^\mu J_\mu = 2 \int d\tau ~n_\mu \nabla_\alpha\left(A^{\alpha\mu} \delta^{(4)}(x-z(\tau)\right) \,, \label{eq:Jn}
\end{equation}
where $A^{\alpha\mu} = u^{[\alpha}p^{\mu]}$. By using the Leibniz rule, we can rewrite this as
\begin{align}
    J_n = 2 \int d\tau\Big[& \nabla_\alpha\left(n_\mu A^{\alpha \mu} \delta^{(4)}\left(x-z(\tau)\right)\right)
    \nonumber \\ & - ~(\nabla_\alpha n_\mu) A^{\alpha \mu} \delta^{(4)}\left(x-z(\tau)\right) \Big] \,. \label{eq:JnAfterLeibniz}
\end{align}
This is as far as we can simplify $J_n$ itself; note that, in the second term, we were already able to take the delta function out of the covariant derivative. For the first term, however, we will have plug equation~\eqref{eq:JnAfterLeibniz} into equation~\eqref{eq:amplitudeBeforeParts}, and integrate by parts in $\mathcal{M}$. In the end, the formula turns out to be
\begin{widetext}
   \begin{equation}
       Z^{\infty,H} = -2\iint \sqrt{-g} ~u^{[\alpha} p^{\mu]} \delta^{(4)}\left(x-z(\tau)\right) \left[(\nabla_\alpha n_\mu + n_\mu \nabla_\alpha) \frac{D_n^\dagger}{\sqrt{-g}} + (\nabla_\alpha \bar{m}_\mu + \bar{m}_\mu \nabla_\alpha) \frac{D_{\bar{m}}^\dagger}{\sqrt{-g}} \right] \mathcal{F}^{\infty, H} e^{i(\omega t' - m\phi')} ~d\tau d^4x' \,.
       \label{eq:amplitudes}
   \end{equation}
\end{widetext}
It is now a straightforward exercise to recover Eq.~\eqref{eq:phi2_inf}.

%
%
\section{Other benchmarks} \label{app:benchmarks}
%

In this section, we thoroughly benchmark our model against several well-known classical results. We begin by comparing it to flat spacetime physics. We cannot naively substitute $M=a=0$ into the formalism developed in Appendix~\ref{app:teuk_method}, as 
our boundary conditions for Teukolsky's radial equation assume the existence of an event horizon. The proper way to solve Teukolsky's equation in this particular setup is detailed in Appendix~\ref{app:flat_formalism}. In the end, we get solutions analogous to the Teukolsky radial functions, related to Bessel functions.

The total power radiated by an electric dipole at rest in flat space is given by the Larmor formula (see Ref.~\cite{Jackson:1998nia}), which is (in units where $4\pi\varepsilon_0 = 1$)
\begin{equation}
    \dot E_L=\frac{1}{3}p_0^2\Omega_{\rm d}^4.
    \label{eq:larmor}
\end{equation}
For a dipole at rest in flat space, the power computed using our formalism matches Larmor's formula~\eqref{eq:larmor} up to machine precision. 

For a dipole in circular motion, the results require a subtler analysis.
Assuming a pointlike source, it is possible to obtain a general expression, within the geometric optics approximation, for the flux measured by stationary observers in a generic stationary spacetime (see Eq.~(S.93) of Ref.~\cite{Santos:2025ass} for the derivation in Kerr). For our system, the energy flux measured by observers at infinity is
\begin{equation}
    \dot E^{\rm d} = -\frac{u_t}{u^t} \left(\frac{1}{3}p_0^2(u^t\Omega_{\rm d})^4\right),
    \label{eq:correctedDipoleFlux}
\end{equation}
where the term $-u_t/u^t$ is related with with the spacetime geometry (note that in flat spacetime $u^t = - u_t = 1$), and $u^t \Omega_{\rm d}$ is the dipole frequency in proper time. Roughly speaking, this formula should be valid when the wavelength is much shorter than the local radius of curvature (i.e. $M\Omega_{\rm d} \gg (M/r_0)^{3/2}$, in Kerr spacetime). Thus, a moving dipole in Minkowski spacetime gets the simple correction to Larmor's formula,
\begin{equation}
    \dot E^{\rm d}=\frac{1}{3}p_0^2(u^t\Omega_{\rm d})^4 = (u^t)^4 \dot E^L.
    \label{eq:flatLarmorCorrected}
\end{equation}

\begin{figure}[h]
    \centering
    \includegraphics[width=1.05\linewidth]{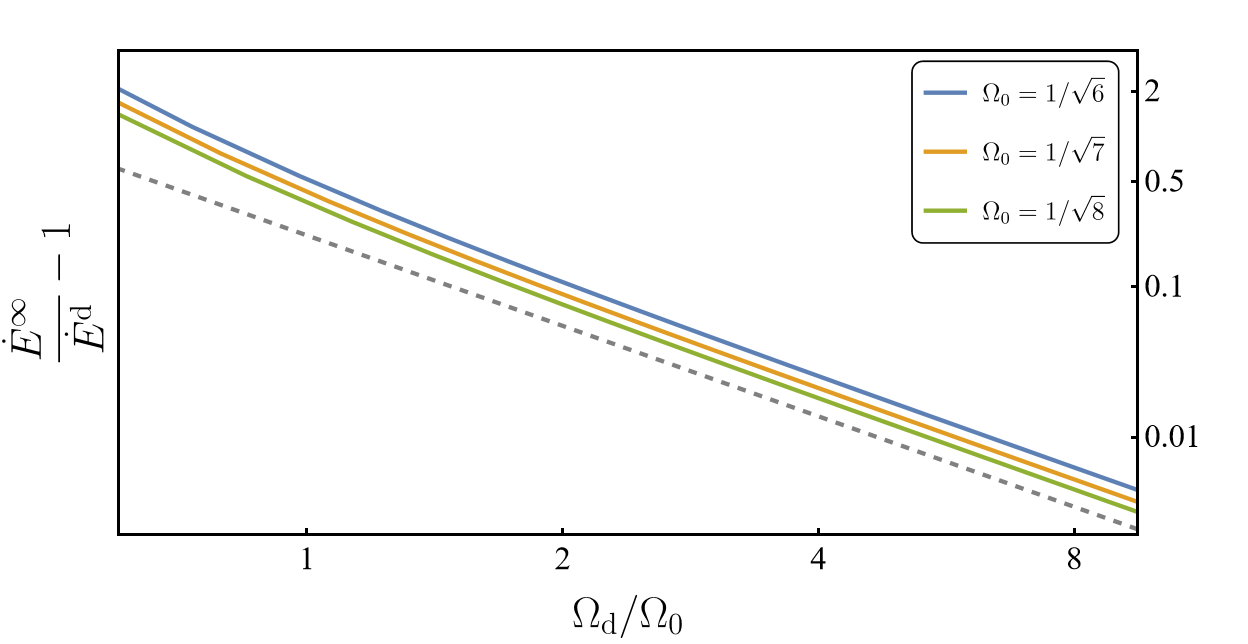}
    \caption{\justifying Normalized difference between computed and expected fluxes at infinity, as a function of dipole frequency $\Omega_{\rm d}/\Omega_0$, for a dipole in uniform circular motion in flat space with frequency $\Omega_0$ and radius $r_0=1$. Eq.~\eqref{eq:correctedDipoleFlux} becomes a good approximation of $\dot E^\infty$ in the $\Omega_{\rm{d}}/\Omega_0\to \infty$ limit, which this logarithmic plot confirms. We also show different lines for different values of $\Omega_0$: as $\Omega_0$ increases, so does the error. This is to be expected, as our approximation works best in the slow-moving regime. The dashed line corresponds to an inverse square law.}
    \label{fig:normDiffFlatSpace}
\end{figure}

We present in Fig.~\ref{fig:normDiffFlatSpace} the normalized difference between the expected power at infinity and the one computed using our formalism, for fixed circular motion with $r_0=1$ and different values of $\Omega_0$, as a function of the frequency $\Omega_{\rm d}$. As the plot shows, the disagreement between the two expressions is large when $\Omega_{\rm d} \sim \Omega_0$, because the circular motion becomes dominant over the oscillatory motion of the dipole. As expected, it becomes increasingly small as $\Omega_{\rm d}$ increases, roughly decreasing as $1/\Omega_{\rm d}^2$.

\begin{figure}[h]
    \centering
    \includegraphics[width=1\linewidth]{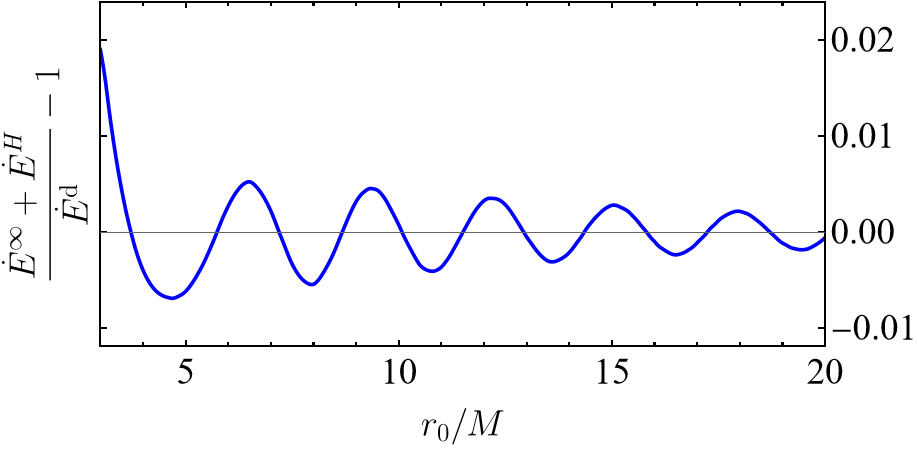}
    \caption{ \justifying Normalized difference between computed (using Eqs.~\eqref{eq:fluxInfty} and \eqref{eq:fluxHorizon}) and expected (using equation~\eqref{eq:correctedDipoleFlux}) total powers, for an electric dipole ($M\Omega_{\rm d} =1$) ar rest in Kerr spacetime ($a=0.2M$), as a function of coordinate radius $r_0$. The accuracy of the approximation is remarkable, with the difference remaining below $1\%$ even as $r_0\sim 4M$, and the oscillations are related to the excitation of quasinormal modes, as shown in Ref.~\cite{Santos:2026lzq}.}
    \label{fig:totalFluxComparison}
\end{figure}

We now move on to curved spacetime. As stated, we expect Eq.~\eqref{eq:correctedDipoleFlux} to be valid when the dipole frequency is much larger than the local radius of curvature. To that end, we consider a dipole at rest in Kerr spacetime ($a=0.2M$), and compare the total flux $\dot E^\infty +\dot E^H$, calculated with Eqs.~\eqref{eq:fluxInfty} and~\eqref{eq:fluxHorizon}, with the predictions of Eq.~\eqref{eq:correctedDipoleFlux}, as a function of the distance to the BH, $r_0$, for $M\Omega_{\rm d}=1$. Our results are depicted in Fig.~\ref{fig:totalFluxComparison}. As we can see, there's a remarkable level of agreement, with the normalized difference between the two quantities remaining below $1\%$ for distances as low as $r_0=4M$.

It is also apparent that there are oscillations in this decay, which are nearly periodic in $r_0$. These are related to the excitation of quasinormal modes of the central BH, as shown in Ref.~\cite{Santos:2026lzq}.

\begin{figure}[h]
\centering
\includegraphics[width=1\linewidth]{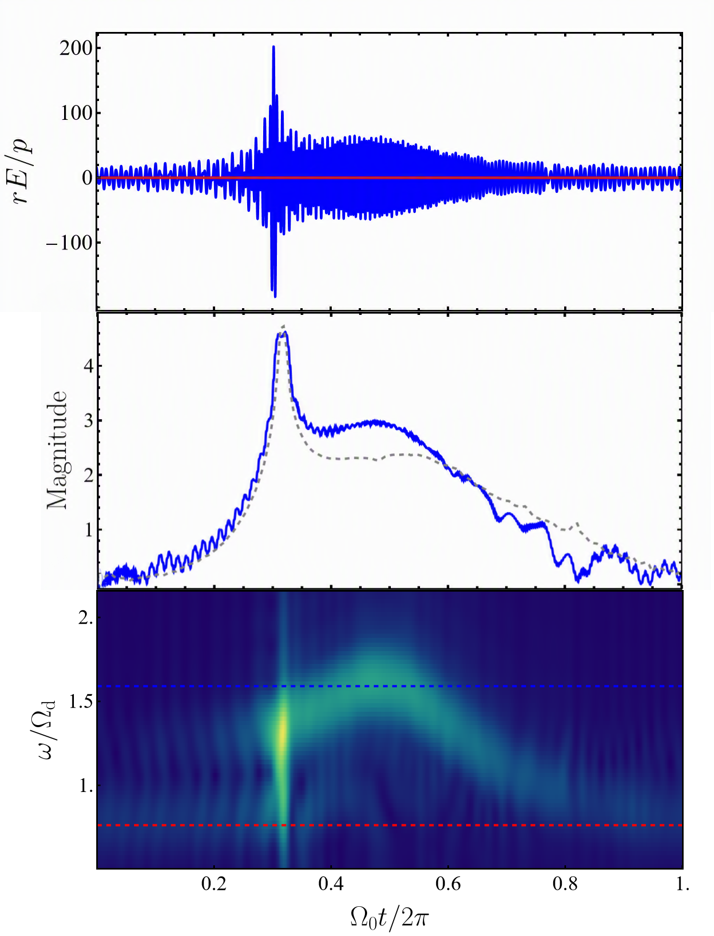}
\caption{\justifying A series of results for our fiducial computation. From top to bottom: waveform seen by an edge-on observer, where blue means polarization in the $\theta$ direction, and red in the $\phi$ direction. \emph{Second:} magnitude of the signal in blue, compared with data from a ray-tracing hotspot simulation in gray. \emph{Third:} spectrogram of the first signal, with blue and red lines corresponding, respectively, to theoretical predictions of blueshift and redshift due to Doppler effect.}
\label{fig:FiducialWaveform}
\end{figure}

Finally, we compare our calculation with an independent numerical treatment of the observed signal, aiming to test whether the waveform itself reproduces the main relativistic effects expected from a moving source in Kerr spacetime, such as Doppler shift, relativistic beaming, and the formation of an Einstein ring. We consider the fiducial set of parameters shown in Table~\ref{tab:FiducialWaveform}, corresponding to a prograde equatorial circular orbit and representative of a dipole emitting low-frequency radio waves, orbiting around a solar-mass BH. The resulting signal is shown in Fig.~\ref{fig:FiducialWaveform}.

\begin{table}[h] \centering \begin{tabular}{c|c|c} \textbf{Quantity} & \textbf{Symbol} & \textbf{Value} \\ Orbital radius & $r_0/M$ & 10 \\ Spin & $a/M$ & 0.7 \\ Inclination & $\iota_d$ & 0 \\ Dipole moment & $p_0$ & 1 \\ Orbital frequency & $M\Omega_0$ & $3.09 \times 10^{-2}$ \\ Precession frequency & $M \Omega_{\rm{p}}$ & 0 \\ Dipole frequency & $M \Omega_{\rm d}$ & 5.0 \end{tabular} \caption{\justifying Parameters for the fiducial computation. The dipole is moving over a prograde circular geodesic around a Kerr BH of mass $M$ and spin $a$.} \label{tab:FiducialWaveform} \end{table}

The waveform is very similar to its gravitational wave counterpart, obtained Ref.~\cite{Santos:2025ass} for an orbiting quadrupole (where the same orbital parameters were used). We also compare our first-principles calculation with the results obtained independently within the geometric-optics approximation, using the ray-tracing software GYOTO \cite{Vincent:2011wz}. We discuss the main features of Fig.~\ref{fig:FiducialWaveform} from top to bottom.

\textbf{First:} the top panel shows the waveform as seen by an edge-on observer. We can already clearly identify signs of Doppler shift (one half of the waveform has a substantially higher frequency than the other) and relativistic beaming (the higher-frequency portion also has a larger amplitude). The frequency and amplitude variations are thus correlated, as expected for a source moving towards and away from the observer.

\textbf{Second:} the second panel provides a more direct comparison with a ray-tracing hotspot simulation. We extract an effective magnitude from our waveform by first computing the squared absolute value of the signal and applying a simple moving average, obtaining a profile $I(t)$. We then express this profile in magnitude units as
\begin{equation}
M(t)=-2.5\log\left(\frac{I(t)}{I_0}\right) \,,
\end{equation}
where $I_0=\min_t I(t)$. Following Ref.~\cite{Santos:2025ass}, we align the peaks of the two profiles before comparing them.

The agreement is remarkably good, particularly around the main peak. This feature can be associated with the formation of an Einstein ring \cite{Hamaus:2008yw,Rosa:2022toh}, providing an independent confirmation of the interpretation proposed in Ref.~\cite{Santos:2025ass}. At later times, however, the hotspot calculation shows a somewhat larger deviation from our result, although the overall agreement remains good. We attribute this discrepancy to the strong interference and beating patterns in the waveform, which make the extraction of a smooth magnitude profile increasingly sensitive to the averaging procedure.

\textbf{Third:} the third panel shows a spectrogram of the signal. The Doppler blueshift and redshift predicted in Ref.~\cite{Santos:2025ass}, represented by the blue and red dashed curves, are clearly reproduced by our first-principles calculation. The spectrogram also shows a peak in intensity coincident with the formation of the Einstein ring.

\section{Formalism for an off-center dipole in flat space} \label{app:flat_formalism}
%
Here we outline how we obtained the results for an off-center, stationary dipole in flat space. The idea is to solve the Teukolsky equation for $s=-1$ and set $M=a=0$. We decompose the field into Fourier-harmonic components, just as in Eq.\eqref{eq:fourierComps}, which yields
\begin{equation}
    \phi_2 \sim \sum_{\substack{\ell, m ,q}} {Z}^N _{\ell m q} \, _{-1}Y _{ \ell m} (\theta) e^{i m \phi} \frac{e^{-i\omega_{00q} u}}{r} \,. \label{eq:psi4_inf_flat}
\end{equation}
Since the dipole is now static, the only frequencies present are $\omega_{00\pm1}$ in Eq.~\eqref{eq:frequencies}, as expected. The spherical symmetry of Minkowski spacetime implies the angular eigenfunctions are now spin-weighted \emph{spherical} harmonics. The amplitudes are determined by
\begin{equation}
    Z_{\ell m q}^N = \sum_ 
    {i= 0} ^{4}\ \sum_ 
    {j = 0} ^{4-i} \mathcal{B}^{(i,j)}_{\ell m q} \frac{d^i }{dr^i} R^{N}\left(r_0 \right) \ \frac{d^j}{d\theta^j}  \bar Y \left( \pi/2\right) \, , \label{eq:amplitudes_flat}
\end{equation} 
where, as in Eq.~\eqref{eq:amplitudes_curved}, $\mathcal{B}$ depends only on the orbital and dipole parameters, $R^{N}\equiv R^{N}_{\ell m d} \equiv R^{N}_{\ell m \omega_{00d}}$ is a solution to the homogeneous radial equation (below), and $\bar Y \equiv {_{-1}\bar{Y}_{\ell m }}$. The homogeneous radial equation is simply
\begin{align}
    & \left(\frac{d^2 }{dr^2}+V(r)\right) \, R^N_{\ell m \omega} = 0 \, , \label{eq:Teukolsky_flat}\\ 
  & V(r) = \omega^2 - \frac{2 i \omega}{r} -\frac{\ell(\ell+1)}{r^2} \, .
\end{align}

We want to find solutions to this equation that are regular at the origin $r=0$. To do so, we first recast the problem into a simpler form using a Chandrasekhar transformation \cite{Chandrasekhar:1985kt}. The latter relates solutions to the Teukolsky equation for $a=0$ (a more generic class than what we are interested in here) to solutions of the Regge-Wheeler equation~\cite{Regge:1957td}, $X_{\ell m \omega}$, which for $M=0$ is just the wave equation 
\begin{align}
    & \left(\frac{d^2}{d r^2} + \omega^2 -\frac{\ell(\ell+1)}{r^2}\right) \, X_{\ell m \omega} =0 \, \label{eq:RW_flat}\, .
\end{align}
Then, to obtain the solution Eq.~\eqref{eq:Teukolsky_flat}, we apply the Chandrasekhar transformation
\begin{equation}
    R^N _{\ell m \omega} = r \left( \frac{d}{dr} + i \omega\right) (X^H _{\ell m \omega}) \, .
\end{equation}
We want solutions to Eq.~\eqref{eq:RW_flat} which are regular at the origin, so we must have 
\begin{equation}
    X^H _{\ell m \omega} (r) = \omega r \,  j_\ell (\omega r) = \eta \, j_\ell (\eta) \, ,
\end{equation}
where we have defined $\eta\coloneq \omega r $, and $j_\ell (\eta)$ are spherical Bessel functions of the first kind. To fix a normalization, we impose that $R^N_{\ell m\omega}$ should have the same behavior at infinity as in \eqref{eq:asymptoticsInfinity}. We obtain
\begin{equation}
    X^{H}_{\ell m\omega}=e^{i\ell\pi/2}r\,j_\ell(\omega r) \,.
\end{equation}
Following this recipe, it easy to obtain the amplitudes using Eq.~\eqref{eq:amplitudes_flat}. Keep in mind, $\mathcal{B}$ is still very complicated because the source term is not simple. To get the full signal, we proceed in the same way as for a dipole in the Kerr geometry: add harmonics until the convergence criterion defined in Ref.~\cite{Santos:2025ass} is satisfied.

\section{Analysis of the face-on signal} \label{app:faceon}

The modulation of the face-on signal is, to a large extent, a kinematic effect rather than a consequence of helicity-dependent scattering by the central BH, as stated in Ref.~\cite{Santos:2025ass}. Since the dipole is boosted, the emitted signal becomes beamed in the direction of motion. As the dipole completes an orbit around the BH, its direction of motion aligns twice with each of the coordinate axes, alternating between the positive and negative directions. This explains why the modulation frequency is twice that of the orbit. We can understand the origin and scaling of this effect using a simple model in flat spacetime (we describe the calculation of the amplitudes in flat space in Appendix~\ref{app:flat_formalism}).

Consider first a source moving with velocity $\beta$ relative to the observer. Let $\Theta$ denote the angle between the observer's line of sight and the boost direction in the source's rest frame, and let $\Theta'$ denote the corresponding angle in the observer's frame. Einstein's aberration formula \cite{Einstein:1905ve} gives
\begin{equation}
\label{eq:aberration}
\cos\Theta'=\frac{\cos\Theta+\beta}{1+\beta\cos\Theta} \,.
\end{equation}
Thus, if the source emits a radiation profile
\begin{equation}
\frac{dN}{d\Theta}=f(\Theta)
\end{equation}
in its rest frame, the corresponding profile in the observer's frame is obtained by changing variables:
\begin{equation}
\frac{dN}{d\Theta'}
=
\frac{d\Theta}{d\Theta'}f(\Theta)
\equiv f'(\Theta') \,.
\end{equation}
Differentiating Eq.~\eqref{eq:aberration} gives
\begin{equation}
f'(\Theta')
=
\gamma(1+\beta\cos\Theta)f(\Theta) \,.
\end{equation}

This two-dimensional description is clearly an idealization of our actual problem, for two reasons: since the source follows an orbit, its velocity changes direction and it is not constant in time, as we are assuming; and the radiation pattern is intrinsically three-dimensional, even though it is toroidal. Nevertheless, the approximation is useful for understanding the face-on signal, since, for sufficiently small $\beta$, acceleration-dependent corrections are subleading, and an observer on the $z$-axis sees radiation in the plane defined by the line of sight and the instantaneous boost direction. We therefore expect this two-dimensional model to capture the leading effect of the source's motion. As we will see, it remains surprisingly accurate even at intermediate velocities.

We now apply this argument to the electric field of the dipole. With our definition of $\Theta$, the angular dependence of the electric field in the source's rest frame is
\begin{equation}
f(\Theta)=\cos\Theta \,.
\end{equation}
Since the electric field is a vector, however, transforming the angular dependence alone is not sufficient: we must also account for the transformation of the angular basis and of the field amplitude. From the aberration relation,
\begin{equation}
\frac{\partial}{\partial\Theta'}
=
\gamma(1+\beta\cos\Theta)
\frac{\partial}{\partial\Theta} \,.
\end{equation}
The vector emission profile in the source frame can be written as
\begin{equation}
\vec E(\Theta)\propto
f(\Theta)\frac{\partial}{\partial\Theta} \,.
\end{equation}
For a face-on observer, $\Theta'=\pi/2$. The aberration formula then implies
\begin{equation}
\cos\Theta=-\beta \implies f(\Theta)=-\beta \,.
\end{equation}
The angular transformation by itself would introduce additional factors of $\gamma$. These factors are, however, compensated by the fact that the dipolar frequency $\Omega_{\rm d}$ is the frequency measured by the observer at infinity, and differs from the frequency as measured in the rest frame of the dipole by a factor of $\gamma$. After accounting for the complete transformation, the leading dependence of the face-on electric-field amplitude is therefore simply
\begin{equation}
\frac{|\vec E_{\beta}|_{\Theta'=\pi/2}}
{|\vec E_{\rm rest}|_{\Theta'=0}}
=\beta \,.
\label{eq:face_on_beta_scaling}
\end{equation}

Equation~\eqref{eq:face_on_beta_scaling} gives a particularly simple prediction: the face-on signal generated by the motion of the source should grow linearly with its orbital velocity. This is verified to be correct and thoroughly explored in Sec.~\ref{subsec:benchmarking}.

\begin{figure}[H]
\centering
\includegraphics[width=1\linewidth]{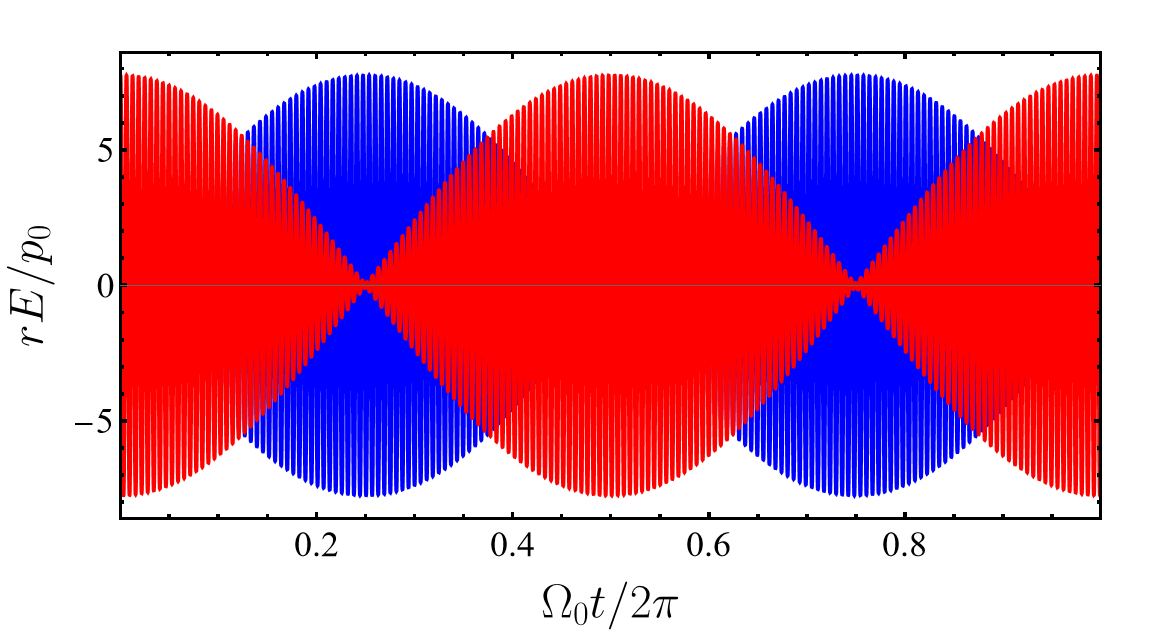}
\caption{\justifying Face-on waveform generated by a dipole in uniform circular motion in flat space, with $r_0\Omega_0=\beta\simeq0.309$ and $M\Omega_{\rm d}=5.0$, matching the setup of Fig.~\ref{fig:face_on}. The blue and red curves show $E_\theta$ and $E_\phi$, respectively. The flat-space waveform closely resembles the full Kerr result in Fig.~\ref{fig:face_on}, with similar amplitudes and overall structure. The main qualitative difference occurs at $\Omega_0t/2\pi=0.25$ and $\Omega_0t/2\pi=0.75$, where the flat-space model predicts $E_\phi=0$, in contrast with the corresponding nonzero values in Fig.~\ref{fig:face_on}.
}
\label{fig:JoaoSuggestion}
\end{figure}


The kinematic contribution is especially important because it is linear in $\beta$, whereas the contribution from scattering off the central BH is in general subdominant. This is true, for instance, for the parameters chosen in Fig.~\ref{fig:face_on}, which are
\begin{equation}
\beta=r_0\Omega_0\simeq0.309 \,,
\qquad
M\Omega_{\rm d}=5.0 \,.
\end{equation}
Since the radiative amplitude scales as $\Omega_{\rm d}^2$, the boosted-source model predicts a characteristic contribution of order
\begin{equation}
\beta(M\Omega_{\rm d})^2
\simeq
0.309\times5.0^2
\simeq 7.7 \,.
\end{equation}

As shown in Fig.~\ref{fig:JoaoSuggestion}, this estimate agrees well with the magnitude of the face-on signal obtained from a flat-space calculation. Comparing with Fig.~\ref{fig:face_on}, we find that, although the resemblance is striking and the amplitudes are comparable, there are also qualitative differences that suggest that beaming cannot be the only effect at play. In particular, our explanation predicts that the envelope of each polarization reaches a minimum of exactly $0$, whereas in the full numerical calculation the minima remain nonzero. This difference suggests that scattering by the black hole contributes to the observed signal, even if it is not the dominant effect. A natural direction for future work is to disentangle these two contributions: the direct signal, arising primarily from relativistic beaming, and the scattered signal, produced by gravitational lensing.

\newpage

\bibliography{References}
\end{document}